\pdfoutput=1

\documentclass[11pt,twoside,a4paper,cmspaper,final,collab]{cms-tdr}
\def\svnVersion{266508}\def\cmsCernNo{2014-063}\def\cmsCernDate{\today}\def\cmsMessage{Submitted to the European Physical Journal C}

\begin{document}\cmsNoteHeader{FSQ-12-026}

\hyphenation{had-ron-i-za-tion}
\hyphenation{cal-or-i-me-ter}
\hyphenation{de-vices}
\RCS$Revision: 250808 $
\RCS$HeadURL: svn+ssh://alverson@svn.cern.ch/reps/tdr2/papers/FSQ-12-026/trunk/FSQ-12-026.tex $
\RCS$Id: FSQ-12-026.tex 250808 2014-07-11 12:22:10Z alverson $
\ifthenelse{\boolean{cms@external}}{}{
\renewcommand{\cmsCollabName}{The CMS and TOTEM Collaborations}
\renewcommand{\cmsNUMBER}{FSQ-12-026}
\renewcommand{\cmsPubBlock}{\begin{tabular}[t]{@{}r@{}l} &CMS \cmsSTYLE\ \cmsNUMBER\\&CERN-PH-EP-TOTEM-2014-002\\\end{tabular}}
\renewcommand{\cmslogo}{\includegraphics[height=2cm]{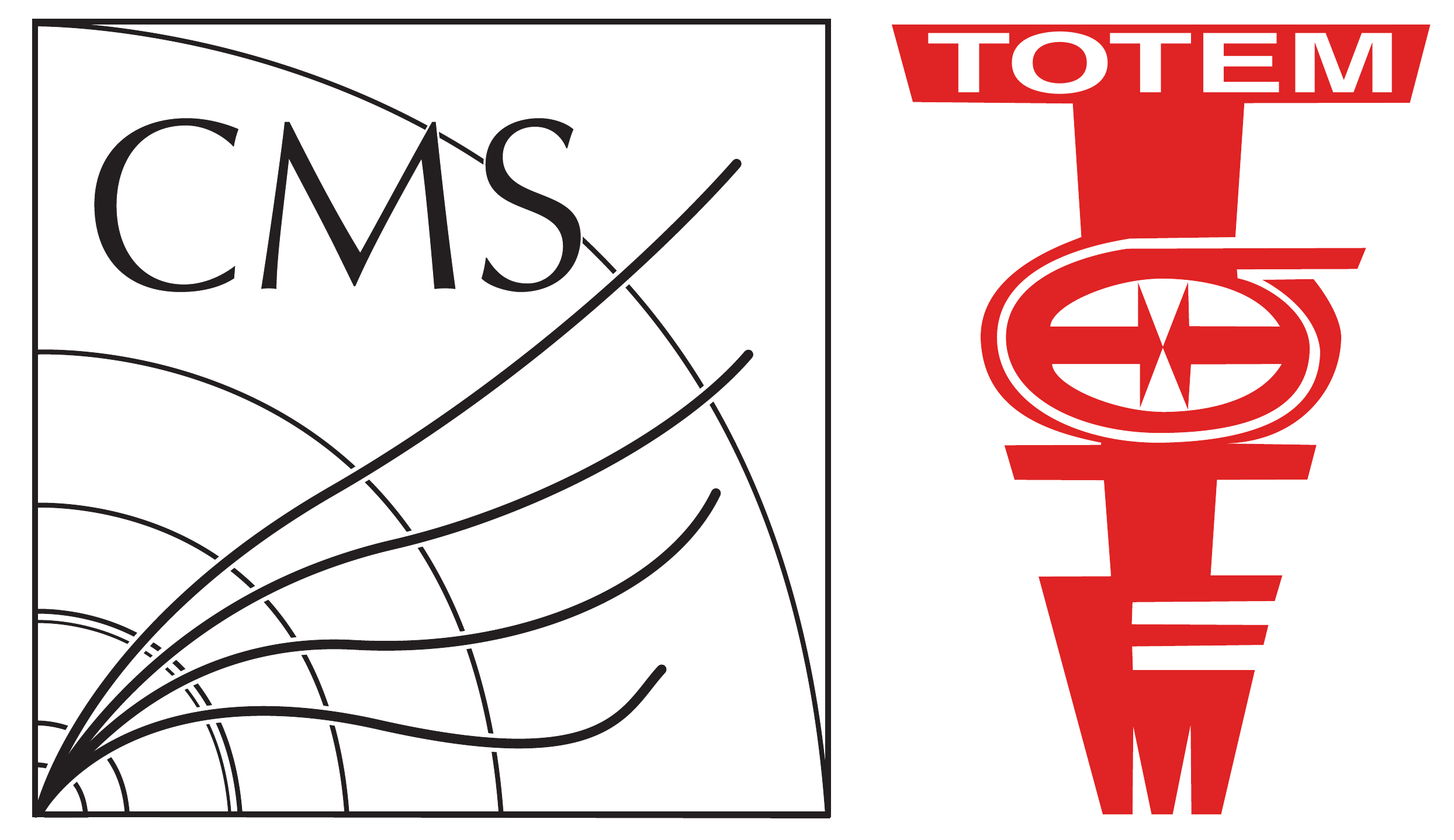}}
\renewcommand{\cmsTag}{CMS-\cmsNUMBER\\&CERN-PH-EP-TOTEM-2014-002\\}
\renewcommand{\appMsg}{See appendices A and B for lists of collaboration members.}
}
\cmsNoteHeader{FSQ-12-026, CERN-PH-EP-DRAFT-TOTEM-2014-002}
\newcommand{\QGS}{\textsc{qgsjet}II\xspace}
\newcommand{\EPOS}{\textsc{epos}\xspace}
\newcommand{\SIBYLL}{\textsc{sibyll}\xspace}
\newlength\cmsFigWidth\ifthenelse{\boolean{cms@external}}{\setlength\cmsFigWidth{0.95\columnwidth}}{\setlength\cmsFigWidth{0.65\textwidth}}
\ifthenelse{\boolean{cms@external}}{\providecommand{\cmsLeft}{Top}}{\providecommand{\cmsLeft}{Left}}
\ifthenelse{\boolean{cms@external}}{\providecommand{\cmsRight}{Bottom}}{\providecommand{\cmsRight}{Right}}
\title{Measurement of pseudorapidity distributions of charged particles in proton-proton collisions at $\sqrt{s} = 8$\TeV by the CMS and TOTEM experiments}

\date{\today}

\abstract{
Pseudorapidity ($\eta$) distributions of charged particles produced in proton-proton collisions at a centre-of-mass energy
of 8\TeV are measured in the ranges $\abs{\eta} < 2.2$ and $5.3 < \abs{\eta} < 6.4$ covered by the CMS and TOTEM detectors, respectively.
The data correspond to an integrated luminosity of $\mathcal{L} = 45$\mubinv.
Measurements are presented for three event categories. The most inclusive category is sensitive to 91--96\% of the total
inelastic proton-proton cross section. The other two categories are disjoint subsets of the inclusive sample that
are either enhanced or depleted in single diffractive dissociation events.
The data are compared to models used to describe high-energy hadronic
interactions. None of the models considered provide a consistent description of the measured distributions.}

\hypersetup{%
pdfauthor={CMS Collaboration},%
pdftitle={Measurement of pseudorapidity distributions of charged particles in proton-proton collisions at sqrt(s) = 8 TeV by the CMS and TOTEM experiments},%
pdfsubject={CMS, TOTEM},%
pdfkeywords={CMS, TOTEM, physics, hadron production}}
\titlerunning{Pseudorapidity measurements in pp at $\sqrt{s}=8\TeV$}
\authorrunning{CMS and TOTEM}
\maketitle
\section{Introduction}
Measurements of the yields and kinematic distributions of particles produced in proton-proton
(pp) collisions at the Large Hadron Collider (LHC) can provide a better understanding of the
mechanisms of hadron production in
high-energy hadronic interactions.
Two types of processes contribute to the production of
most of the final-state particles at LHC energies. Semi-hard (multi)parton scattering, with
exchanged momenta of a few \GeV,
is the dominant contribution. Diffractive scattering in more
peripheral pp interactions, where one or both protons survive the interaction and/or are excited
into a low-mass state, accounts for 15--40\%~\cite{Ostapchenko_2011nk, ALICE2013_xsec} of the pp inelastic cross section.
As the particle multiplicity produced in these processes is modeled phenomenologically
in the existing Monte Carlo (MC) event generators of
hadronic interactions, experimental results provide an important input for tuning of the
models.

The results presented here focus on the charged-particle multiplicity density ($\rd{}N_{\text{ch}}/\rd{}\eta$,
also referred to as the pseudorapidity distribution)
corrected down to zero transverse momentum (\pt), in the pseudorapidity ranges $\abs{\eta} < 2.2$ and
$5.3<\abs{\eta}<6.4$, where $\eta$ is defined as $-\ln[\tan(\theta/2)]$, with $\theta$ being the polar
angle of the particle trajectory with respect to the anticlockwise-beam direction.
Inclusive measurements of the $\eta$ and
\pt distributions of charged particles have previously been performed in pp and $\Pp\Pap$   collisions for different
centre-of-mass energies and phase space regions~\cite{CMSdndeta1, CMSdndeta2, ATLAS2011, ALICE2010a, ALICE2010, LHCb2011, TOTEM2012_dndeta, CDF:2009, CDF:1990, UA1, UA4_dndeta, UA5}.

In this paper, the data samples were collected with a minimum bias trigger generated by at least one arm of the TOTEM
T2 telescopes, which also triggered the readout of the Compact Muon Solenoid (CMS).
Three event samples with different final state topologies are selected offline: a sample of inclusive inelastic pp events,
a sample dominated by non single diffractive dissociation (NSD) events, and a sample enriched in single diffractive
dissociation (SD) events.
The measured data are compared to the predictions of MC event generators that model
pp collider data and high-energy cosmic ray hadronic interactions.

\section{Experimental setup}
The CMS and TOTEM experiments use a right-handed coordinate system, with the origin at the nominal interaction point (IP),
the $x$-axis pointing to the centre of the LHC ring, the $y$-axis pointing upwards, and the $z$-axis pointing along the
anticlockwise-beam direction. The azimuthal angle, $\phi$, is measured in the $(x,y)$ plane, where $\phi=0$ is the  $+x$ and $\phi=\pi/2$ is the $+y$ direction.

A complete description of the CMS detector can be found in Ref.~\cite{CMSdetector}.
The central feature of the CMS apparatus is a superconducting solenoid of 6\unit{m} internal diameter, providing a uniform magnetic field of 3.8\unit{T} parallel to the beam axis. Inside the magnetic field are the pixel tracker, the silicon-strip tracker,
the lead tungstate electromagnetic calorimeter, and the brass/scintillator hadron calorimeter. Muons are measured in gas-ionisation
detectors embedded in the steel return yoke outside the solenoid. In addition to the barrel
and endcap detectors, which extend up to $\abs{\eta}=3.0$, the steel/quartz-fibre hadron forward calorimeters (HF)
cover the region $2.9<\abs{\eta}<5.2$.
The tracking detector consists of 1440 silicon pixel and 15\,148 silicon strip detector modules. The barrel is composed of
3 pixel and 10 strip  layers around the interaction point at radii from 4.4 to 110\unit{cm}. The forward and backward
endcaps each consist of 2 pixel disks and 12 strip disks in up to 9 rings. Three of the strip
rings and four of the barrel strip layers contain an additional plane, with a stereo angle of 100\unit{mrad}, to provide
a measurement of the $r$-coordinate and $z$-coordinate, respectively.
The tracker is designed to provide a longitudinal and transverse impact parameter resolution of about 100\mum and a \pt resolution of about 0.7\% for 1\GeVc charged particles at $\eta=0$~\cite{tracking}.

The TOTEM experiment~\cite{totem, totemPerformance:2013} is composed of three subdetectors: the Roman pots, and the T1 and T2 telescopes.
Minimum bias events are triggered by the two T2 telescopes, which are placed symmetrically on each side
of the IP at about $|z|=$14\unit{m}. They detect charged particles produced in the polar angular range of $\approx$3--10\unit{mrad} ($5.3<\abs{\eta}<6.5$), with full azimuthal acceptance.
Each telescope consists of two half-arms, with each half-arm composed of
10 semicircular planes of triple-GEM (gas electron multiplier) chambers, arranged within a 40\unit{cm} space along the $z$-axis.
Each chamber provides two-dimensional information on the track position covering 192$^{\circ}$ in azimuthal angle
with a small vertical overlap region between chambers of two neighbouring half-arms.
Every chamber has a double-layered readout board containing two columns of 256 concentric strips
(400\mum pitch, 80\mum width) to measure the radial coordinate and a matrix of 1560 pads, each covering
$\Delta\eta\times\Delta\phi\approx0.06\times0.018\unit{rad}$, to measure the azimuthal coordinate, and for triggering.
The radial and azimuthal coordinate resolutions are about 110\mum and 1$^{\circ}$, respectively~\cite{NimT2}.
Local angles of single tracks are reconstructed with an average resolution of 0.5\unit{mrad}, and the track pseudorapidity resolution
for charged particles is better than 0.05~\cite{MirkoTesi}, once the track is identified as coming from the vertex.

The detailed MC simulations of the CMS and TOTEM detectors are based on \GEANTfour~\cite{Geant}. Simulated events are
processed and reconstructed in the same manner as collision data.

\section{Monte Carlo models}
{\tolerance=1000
Various MC event generators for hadronic collisions are used for data corrections and for comparison with the final, fully corrected results.
The \PYTHIA{}6 (version 6.426) \cite{sjostrand2006} generator is used with tune Z2* and \PYTHIA{}8 (version 8.153) \cite{sjostrand2008}
with tune 4C \cite{tune4C_2011}.
These programs provide different descriptions of the diffractive component and they both use a model \cite{skands2007} in which multiple
partonic interactions
are interleaved with parton showering. The Z2* tune \cite{Chatrchyan:2011wm} uses the CTEQ6L1 \cite{cteq6L_2002}
parton distribution function (PDF) set. Tune 4C of \PYTHIA{}8 is based on early LHC data \cite{tune4C_2011}. Parton showers in \PYTHIA are
modeled according to the Dokshitzer--Gribov--Lipatov--Altarelli--Parisi (DGLAP) \cite{Gribov:1972ri, Altarelli:1977zs, Dokshitzer:1977sg} prescription and hadronisation is based on the Lund string fragmentation model \cite{andersson:1983}.
Diffractive cross sections are described by the
Schuler--Sj{\"o}strand model \cite{Sch:94}. In \PYTHIA{}6, particle  production from a low-mass state, $X$, with $M_X < 1$\GeVcc is
treated as an isotropic two-body decay, while for high-mass states it is based on the string model. In \PYTHIA{}8, the same model
is used to generate the cross section and the diffractive mass, but particle production is modeled differently.
For low-mass states, the string model is used, but
for higher masses ($M_X > 10$\GeVcc) a perturbative description of pomeron-proton scattering is introduced, based on diffractive
PDFs \cite{H12006, H12007, diffractivePDFs_2013}, which represent probability distributions
for partons in the proton under the constraint that the proton emerges intact from the collision.
The non-perturbative string model introduces a mass dependence on the relative probability for a pomeron to couple to a quark or a gluon \cite{navin}.
The perturbative treatment of pomeron-proton scattering results in harder \pt spectra and higher multiplicity for
diffractive events generated
with \PYTHIA{}8 than for those obtained with \PYTHIA{}6.
\par}

The \HERWIGpp (version 2.5.0)~\cite{bahr:2008} MC event generator, with a recent tune to LHC data (UE-EE-3
with CTEQ6L1 PDFs~\cite{cteq6L_2002, Gieseke_2012ft}),
is also used for comparison with the data.
This generator is based on matrix element calculations similar to those used in \PYTHIA. However, \HERWIGpp
features DGLAP based parton showers ordered in angle and uses cluster
fragmentation for the hadronisation~\cite{gieseke_2011}.
The description of hard diffractive processes also makes use of diffractive PDFs; however soft diffraction is not implemented.

The data are also compared to predictions from two MC event generators used in cosmic ray physics \cite{denterria:2011}:
\EPOS \cite{werner2006} with the LHC tune (based on \EPOS 1.99) \cite{Pierog_2013_eposLHC} and \QGS-04 \cite{ostapchenko2011}.
Both models include contributions from soft- and hard-parton dynamics. The soft component is described in terms of the exchange
of virtual quasi-particle states, as in Gribov's Reggeon field theory \cite{gribov1968}, with multi-pomeron exchanges
accounting for underlying-event effects. At higher energies, the interaction is described in terms of the same degrees of
freedom, but generalised to include hard processes via hard-Pomeron scattering diagrams, which are
equivalent to a leading-order perturbative Quantum Chromodynamics (QCD) approach with DGLAP evolution.
These models are retuned to LHC data \cite{Pierog:2013}, including cross section measurements by TOTEM,
and charged-particle multiplicity measurements in the central region, by ALICE and CMS, at $\sqrt{s}=7$\TeV.

\section{Datasets}
The data were collected in July 2012 during a dedicated run with low probability ($\sim$4\%) of overlapping pp interactions in the same bunch
crossing (pileup) and a non-standard $\beta^*$ = 90\unit{m} optics configuration, where $\beta^*$ is the amplitude function of the beam at the interaction point.
These data correspond to an integrated luminosity of $\mathcal{L} = 45\mubinv$.
A minimum bias trigger was provided by the TOTEM T2 telescopes
and contributed to the CMS global trigger decision, which initiated simultaneous readout of both the CMS and TOTEM detectors.
The CMS orbit-counter reset signal delivered to the TOTEM electronics at the start of the run assures the time synchronisation of
the two experiments.
Events are combined offline by requiring
that both the CMS and TOTEM reconstructed events have the same LHC orbit and bunch numbers.
The minimum bias trigger required at least one track candidate (trigger track) in the T2 detector,
in either $z$-direction~\cite{TOTEM_Inel}.
With this selection the visible cross section seen by T2 has been estimated to be 91--96\% of the total pp inelastic
cross section at $\sqrt{s}=8$\TeV~\cite{TOTEM_total8TeV}.
Zero bias data, obtained by triggering on random bunch crossings, are used to measure the trigger efficiency
and to cross-check the pileup probability estimate.

MC samples were used to determine the event selection efficiency and the tracking performance.
The efficiency corrections and related uncertainties for the CMS tracker are based on the \PYTHIA{}6 and \EPOS samples.
The MC-based corrections for the TOTEM T2 detector and the corresponding uncertainties were determined by using \PYTHIA{}8 and \EPOS,
which were found to bracket the measured $\rd{}N_{\text{ch}}/\rd{}\eta$ distributions in the forward region.

\section{Event selection and track reconstruction}
\label{sec:Event_selection}
The T2 track reconstruction is based on a Kalman filter-like algorithm,
simplified thanks to the small amount of material in the GEM planes and
the weak magnetic field in the T2 region~\cite{MirkoTesi}.
In these conditions, the particle trajectory can be successfully reconstructed with a straight-line fit.
Single tracks are reconstructed with almost 100\% efficiency for $\pt>20\MeVc$, but because of multiple scattering and the magnetic field,
tracks can be identified as coming from the IP with an efficiency that increases as a function of \pt and is
greater than 80\% for $\pt>40\MeVc$~\cite{totemPerformance:2013}.
The pseudorapidity of a track in T2 is defined as the average pseudorapidity of all T2 track hits,
calculated from the angle between the $z$-axis and the line joining the hit and the nominal IP\@.
This definition is adopted on the basis of MC simulation studies and gives an
optimal estimation of the pseudorapidity of the selected primary (\ie produced at the IP) particle.
Because of the small scattering angle of the particles reconstructed in T2, the position of the vertex does not
affect significantly the reconstruction of the track pseudorapidity.
Due to the limited number of primary particles in T2 per event, no vertex reconstruction based on T2 information
is used for the analysis.
The pseudorapidity region $5.4<\abs{\eta}<5.6$ is not included in the analysis because of the large effect that the beam pipe cone
at $\abs{\eta}\approx 5.5$ has on the propagation of the primary particles.

\begin{figure}[htb!]
\centering
\includegraphics[width=\cmsFigWidth]{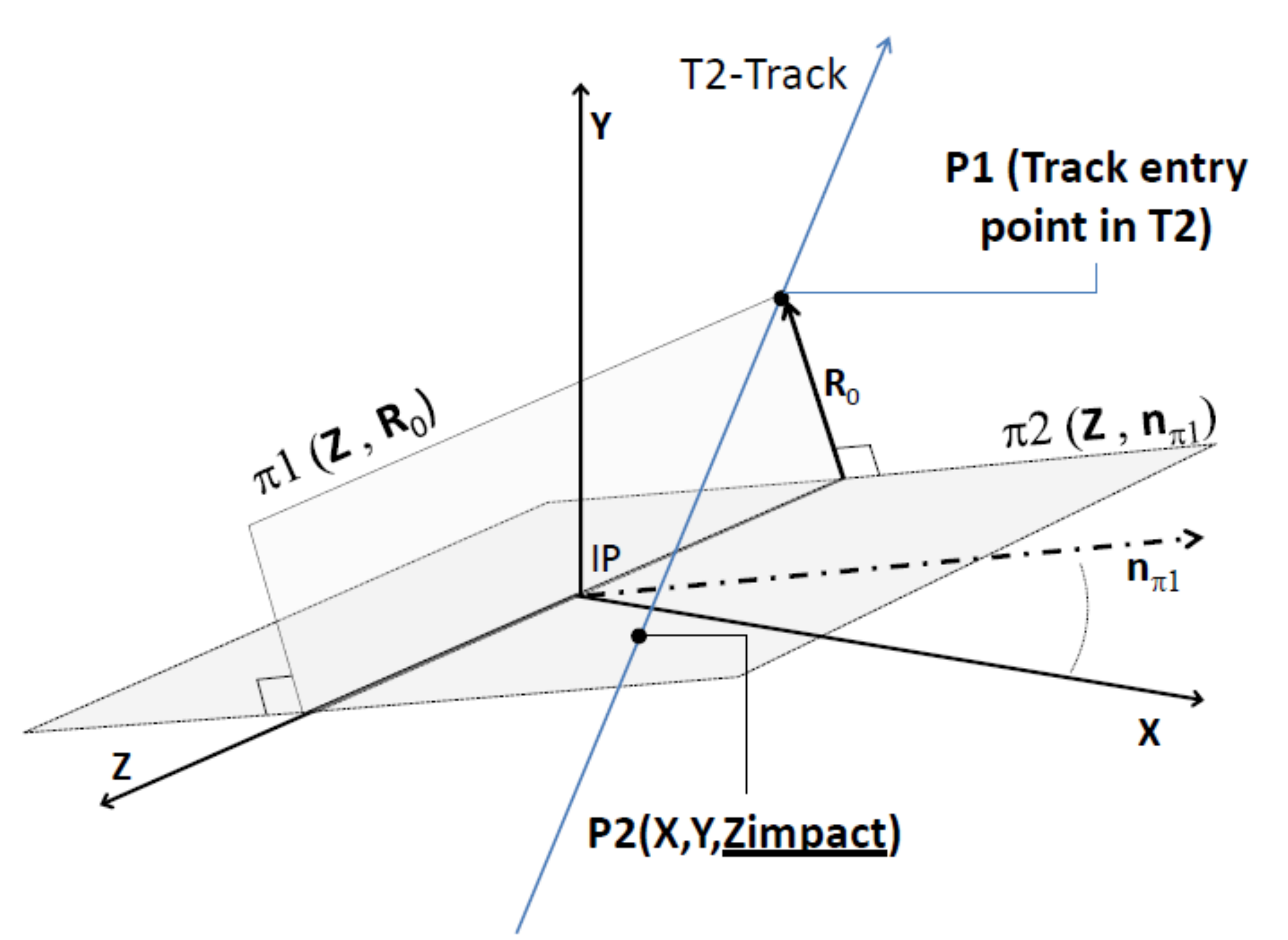}
\caption{Definition of the $z_\text{impact}$ parameter.}
\label{fig:zimpact}
\end{figure}

About 80\% of the reconstructed tracks in T2 are due to non-primary particles, hereafter referred to as secondary particles,
that are mainly electrons and positrons generated
by photon conversions at the edge of the HF calorimeter of CMS
and in the conical section of the beam pipe at $\abs{\eta}\approx5.5$. It is therefore important to discriminate these particles from the primary
charged particles. The most effective primary/secondary particle separation is achieved by using the $z_\text{impact}$ track parameter
(see Fig.~\ref{fig:zimpact}), which is defined as the $z$ coordinate of the
intersection point between the track and the plane ``$\pi$2''. This is the plane which contains the $z$-axis and is orthogonal
to the plane ``$\pi$1'' defined by the $z$-axis and containing the
track entry point in T2~\cite{TOTEM2012_dndeta}. This parameter is found to be stable against residual misalignment biases.
Simulation studies demonstrate that the $z_\text{impact}$ distribution in the primary-particle region can be described by the sum of
two Gaussian distributions plus an exponential distribution. The Gaussians, hereafter referred to as a ``double-Gaussian''
distribution, are mainly due to primary particles, whereas the exponential distribution accounts for most secondary particles.
Simulations predict a contamination
of the double-Gaussian distribution by secondary particles of about 20\%, primarily given by photons converted in the material between
the IP and T2, with a contribution from decay products of strange particles; the $z_\text{impact}$ distribution of these particles
is a Gaussian centred around $z_\text{impact}=0$.

\begin{figure}[htb!]
\centering
\includegraphics[width=\cmsFigWidth]{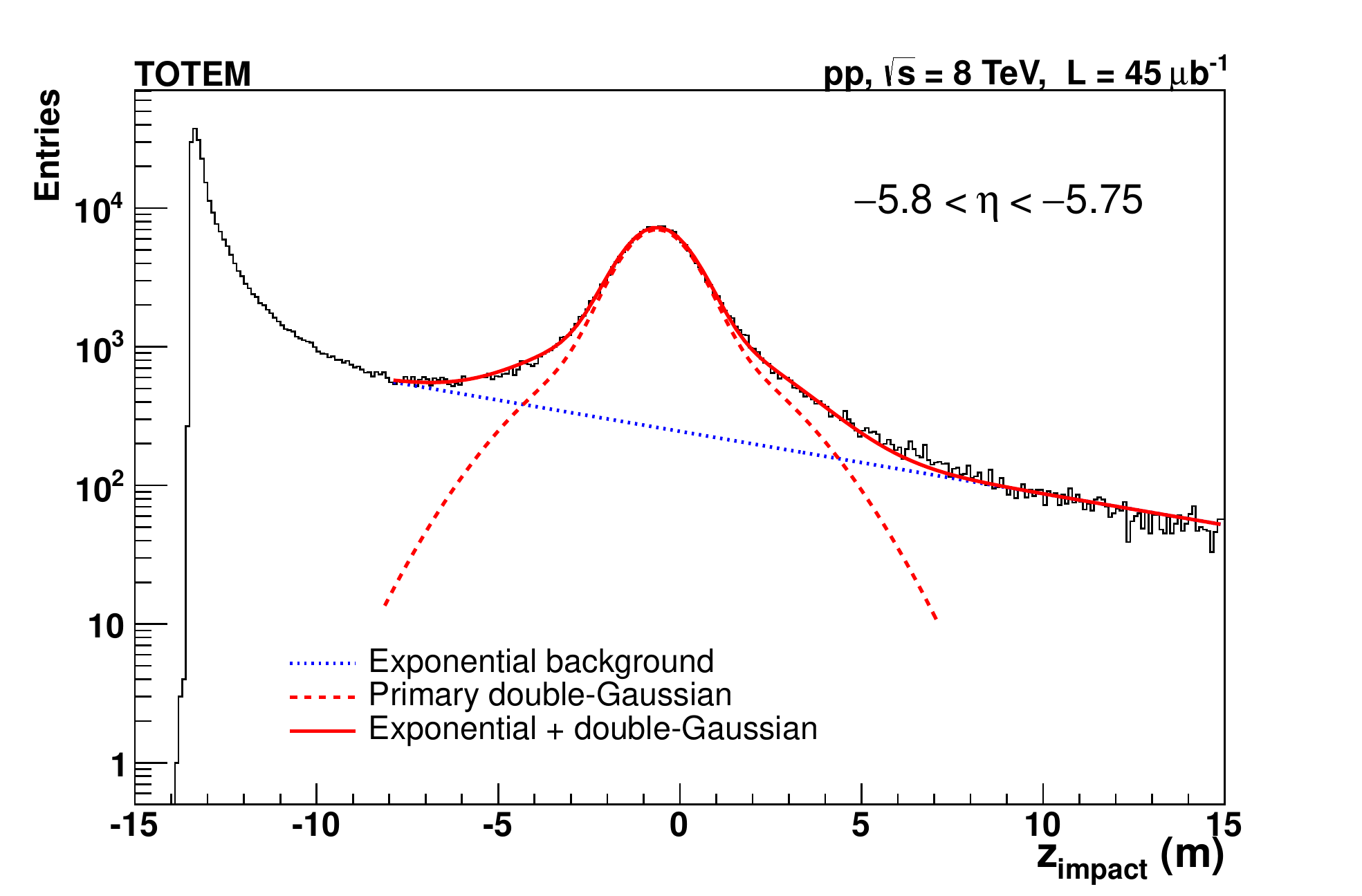}
\caption{The $z_\text{impact}$ parameter distribution measured in the data, for tracks reconstructed in one T2 half-arm in the range $-5.8<\eta<-5.75$.
A global (double-Gaussian + exponential function) fit, performed in the range from $-8$ to 15\unit{m} is shown by the solid curve.
The dotted curve represents the exponential component from secondary particles, while the dashed curve is the double-Gaussian component,
mainly due to primary tracks.
}
\label{fig:MCPrimSecfitsBeforeAfterb}
\end{figure}

Figure~\ref{fig:MCPrimSecfitsBeforeAfterb} shows the $z_\text{impact}$ distribution at the median $\eta$
of the inclusive forward pseudorapidity distribution. A combined fit is performed for each $\eta$ bin of the $\rd{}N_{\text{ch}}/\rd\eta$ distribution
with the sum of a double-Gaussian and an exponential function, yielding standard deviations
(amplitudes) of both Gaussian functions that increase (decrease) with $\eta$.  The mean, required
to be the same for both Gaussian distributions, the standard deviations, and the amplitudes of the two Gaussian functions,
as  well as the mean and the amplitude of the exponential, are left free in the fit.
The widths of the double-Gaussian distributions are consistent with
the observed angular resolution of about 0.5\unit{mrad} for the T2 track reconstruction.
The relative abundance  of secondary particles is found to be smaller for higher $\abs{\eta}$.
The fit of the $z_\text{impact}$ distribution is also repeated by using a second degree polynomial for the description of the background.
The results are found to be stable with respect to the choice of the background function.
The integral of the fitting function approximates the area of the $z_{\text{impact}}$ distribution to within 1\%.

The T2 tracks are considered ``primary candidates'' if they are in the $z_\text{impact}$ range corresponding to 96\%
of the area of the double-Gaussian, taken symmetrically around the mean.

The standard CMS track reconstruction algorithm is based on a combinatorial track finder (CTF)~\cite{tracking}. The collection of reconstructed tracks
is produced by multiple iterations of the CTF track reconstruction sequence, in a process called iterative tracking.
The reconstruction of the interaction vertices in the event uses the available reconstructed track collection.
Prompt tracks are selected based on given quality criteria, and the selected tracks are then clustered in $z$ using
a ``deterministic annealing'' (DA) algorithm~\cite{DAalgo}. After identifying candidate vertices based on the DA clustering,
the candidates containing at least two tracks are fitted by means of an
adaptive vertex fit, where tracks in the vertex are assigned a weight between 0 and 1 based
on their compatibility with the common vertex.
In the central region, covered by the CMS tracker, high-purity primary tracks~\cite{TRK-10-001} are selected with
$\pt>0.1$\GeVc and
relative \pt uncertainty less than 10\%.
To maximise the track-vertex association efficiency, a selection is
applied on the track impact parameter, both along the $z$-axis and in the
transverse plane. The impact parameter with respect to the beam spot in
the transverse plane, $d_{xy}$, is required to be $\abs{d_{xy}/\sigma_{xy}} < 3$,
while for the point of closest approach to the primary vertex along the $z$-direction, $d_{z}$,
the requirement $\abs{d_z/\sigma_z} < 3$ is imposed. Here
$\sigma_{xy}$ and $\sigma_z$ denote the uncertainties in $d_{xy}$ and $d_z$, respectively.
The analysis is restricted to $\abs{\eta}<2.2$, to avoid effects from tracks
close to the geometric edge of the tracker. Events with more than one reconstructed vertex are discarded,
thus reducing the effect of pileup to a negligible level ($<1$\%).

The pseudorapidity distributions of the charged particles are measured in the central and forward regions for three different event samples,
with topologies corresponding to three different event selection criteria. An inclusive sample of events is selected by requiring at
least one primary track candidate in T2. Event samples enhanced in non single diffractive dissociation (NSD) and
single diffractive dissociation (SD) events are also selected, the former defined by requiring
a least one primary candidate in each of the two T2 telescopes and the latter by selecting events with at least one primary
candidate in one T2 telescope and none in the other. Therefore, the intersection of the NSD-enhanced and SD-enhanced samples is empty,
while the union is the inclusive sample.

The inclusive sample includes $\sim$99\% of non-diffractive events. The reconstruction efficiency for diffractive
events is 50\% for a diffractive mass $M\sim 3.6$\GeVcc and increases rapidly to 99\% for $M>10$\GeVcc.
Most of the non-diffractive and double diffractive events, as well as the single diffractive events with masses
larger than 1.3\TeVcc, produce particles in both T2 telescopes and are therefore included in the NSD-enhanced sample.
Simulation studies based on \PYTHIA{}8 and \EPOS show that the fraction of NSD events in the SD-enhanced sample amounts to 45--65\%.

\section{Data analysis}
The pseudorapidity density measurements presented here refer to ``stable'' primary charged particles, with an average lifetime
longer than $3\times10^{-11}$ s, either directly produced in pp collisions or from decays of particles with
shorter lifetimes.
Such a definition, consistent with that of previous studies~\cite{CMSdndeta1, CMSdndeta2, ATLAS2011, ALICE2010a, ALICE2010, TOTEM2012_dndeta},
considers as secondary particles the decay products of \PKzS~ and~\PgL~hadrons and all of the charged particles generated by interactions
with the material in front and around the detectors.

\subsection{Trigger efficiency correction}
\label{sec:Trigger_dndeta}
\begin{figure}[htb!]
\centering
\includegraphics[width=\cmsFigWidth]{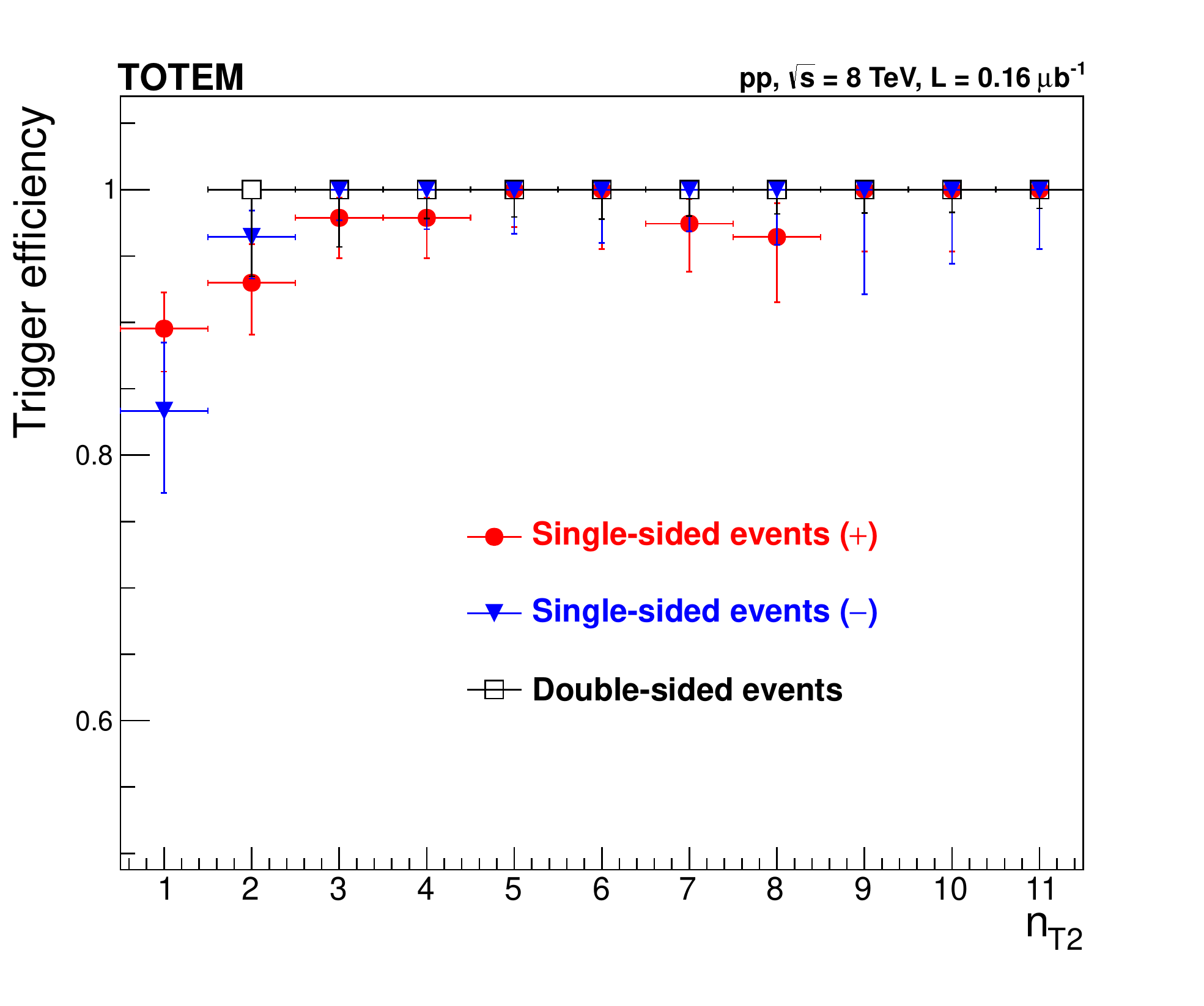}
\caption{Trigger efficiency as a function of the total track multiplicity in the T2 telescopes for single-sided events with primary candidates in only the $z>0$ ($+$) or $z<0$ ($-$) telescope and for double-sided events with primary candidates in both telescopes.}
\label{fig:Treff}
\end{figure}

The trigger efficiency is determined with a zero bias event sample separately for
events with primary track candidates reconstructed offline in both arms of T2 and in only one arm.
All zero bias data taken were used to determine the trigger inefficiency.
The inefficiency of the trigger is mainly due to non-operating and noisy channels.
For each event category, the trigger efficiency is calculated as a function of the total number of tracks reconstructed offline
in T2, $n_{\text{T2}}$,
as $\epsilon_\text{{trig}}$ = $N(n_{\text{T2}})_\text{trig}/N(n_{\text{T2}})_\mathrm{zb}$,
where $N(n_{\text{T2}})_\text{trig}$ is the number of events with the total T2 track multiplicity $n_{\text{T2}}$ passing the trigger selection and $N(n_{\text{T2}})_\mathrm{zb}$ is the number of events with $n_{\text{T2}}$ tracks selected with the zero bias trigger.
The measured trigger efficiency is shown in Fig.~\ref{fig:Treff}.

The trigger efficiency correction, $1/\epsilon_\text{{trig}}$, is applied separately for the three event categories and is significant
for events with $n_{\text{T2}}=1$, while it approaches unity for $n_{\text{T2}}\ge 3$.
The overall trigger inefficiency results in a 0.1--0.2\% relative correction to the total $\rd{}N_\text{ch}/\rd\eta$ distributions
for the three event categories.
The value of the associated systematic uncertainty is conservatively assumed to be equal to this relative correction.

\subsection{Event selection correction}
\label{sec:EventSelCorr_dndeta}
In order to take into account the differences between the sample defined at the MC-particle level (``gen selected'')
and the one selected based on the reconstructed tracks (``reco selected''), a correction
factor needs to be introduced.
This correction is calculated for each $\eta$ bin and event category from the ratio
\begin{equation}
C_{\text{sel}}(\eta)=\frac{\rd{}N_{\text{ch}}/\rd\eta_{\text{gen}}|_{\text{gen selected}}}{\rd{}N_{\text{ch}}/\rd{}\eta_{\text{gen}}|_{\text{reco~selected}}},
\end{equation}
where the numerator is the pseudorapidity density obtained from the MC simulation for events selected based on
the charged particles within the T2 acceptance, while the denominator is the
density obtained by selecting the simulated events according to the topology of
the primary candidates in T2, as explained in Section 5.
In general, $C_{\text{sel}}$ is different from unity because of migrations between the different event categories.
Because of misidentification of secondary particles as primaries and of track reconstruction inefficiencies, an event can be classified (according to the configuration of its reconstructed tracks) in a category that does not reflect its true
charged-particle content.

For the inclusive and NSD-enhanced analysis, $C_{\text{sel}}$ is evaluated with \PYTHIA{}8 and \EPOS.
Moreover, to quantify possible biases related to this correction, the analyses are repeated with the same selection method defined in Section~\ref{sec:Event_selection} but without the primary candidate $z_{\text{impact}}$ requirement.

As the SD-enhanced multiplicity is smaller than the NSD-enhanced multiplicity, a larger selection inefficiency is expected
for the former class of events. Moreover, as the SD-enhanced sample represents only 26\% of the inclusive sample, the NSD events
that are wrongly identified as SD can introduce a large bias in the measured SD-enhanced $\rd{}N_\text{ch}/\rd{}\eta$
distributions. Additional studies of the  event selection strategy for the SD-enhanced analysis are therefore performed.
The analysis is repeated with different event selection strategies and $C_{\text{sel}}$ is reevaluated for each as a function of $\eta$ and of the track multiplicity in T2.
The selection methods differ in the maximum number of tracks and of primary candidates reconstructed in T2 on each side of the IP\@.
Simulation studies show that, depending on the method and the MC generator used, the selection
efficiency is in the range of 70--90\%. The purity of the SD-enhanced sample, defined
as the fraction of the selected events with primary charged particles
in only one arm of T2, varies between 66\% and 81\%. The dependence of the SD-enhanced $\rd{}N_\text{ch}/\rd{}\eta$ distributions
on the selection methods is used to evaluate the systematic uncertainty related to the event selection.
More details on the numerical values of $C_{\text{sel}}$ are given in Sections~\ref{sec:CMS_dndeta} and \ref{sec:T2_dndeta}.

\subsection{Measurement of \texorpdfstring{$\rd{}N_{\text{ch}}/\rd{}\eta$}{dN[ch]/d eta} in the central region}
\label{sec:CMS_dndeta}
The charged particle pseudorapidity distributions in the central region are obtained
from the raw distributions of charged tracks
after applying a number of corrections according to the formula:
\ifthenelse{\boolean{cms@external}}{
\begin{multline}
\frac{\rd{} N_\text{{ch}}}{\rd{} \eta} = \\ \frac{ C_{\text{sel}}(\eta)\sum_{\text{evt}}\omega_\text{{evt}}(n_{\text{CMS}}, n_{\text{T2}})\sum_{\text{trk} \in S}\omega_\text{{trk}}(n_{\text{CMS}},\pt,\eta ) }{\Delta \eta
\sum_{n_{\text{CMS}}} N_\text{{evt}}(n_{\text{CMS}},n_{\text{T2}})\omega_\text{{evt}}(n_{\text{CMS}}, n_{\text{T2}})},
\label{eq:dnde}
\end{multline}
}{
\begin{equation}
\frac{\rd{} N_\text{{ch}}}{\rd{} \eta} =  \frac{ C_{\text{sel}}(\eta)\sum_{\text{evt}}\omega_\text{{evt}}(n_{\text{CMS}}, n_{\text{T2}})\sum_{\text{trk} \in S}\omega_\text{{trk}}(n_{\text{CMS}}, \pt,\eta ) }{\Delta \eta
\sum_{n_{\text{CMS}}} N_\text{{evt}}(n_{\text{CMS}},n_{\text{T2}})\omega_\text{{evt}}(n_{\text{CMS}}, n_{\text{T2}})},
\label{eq:dnde}
\end{equation}
}
where $S$ is the sample of tracks that pass the selection criteria for a given $\eta$ bin,
$n_{\text{CMS}}$ and $n_{\text{T2}}$ is the total number of reconstructed tracks in the CMS tracker and T2, respectively, $N_\text{{evt}}$ is the number of triggered events in the corresponding
track multiplicity bins, $\omega_\text{{evt}}$ corrects for the trigger and the vertex
reconstruction efficiencies,
$\omega_\text{{trk}}$ corrects for the tracking efficiency and the effect of non-primary tracks, and $C_{\text{sel}}$
corrects for the effect of the event and primary track selection with T2.
The bin width in $\eta$ is $\Delta\eta=0.2$.

The event correction, $\omega_\text{{evt}}$, depends on the track multiplicity in T2, $n_{\text{T2}}$, as well as on
the multiplicity in the CMS tracker because of the minimum number of
tracks required in the vertex reconstruction. It is given by:
\begin{equation}
 \omega_\text{{evt}}(n_{\text{CMS}},n_{\text{T2}})  = \frac{1}{\epsilon_\text{{trig}}(n_{\text{T2}})\epsilon_{\text{vtx}}(n_{\text{CMS}})},
\label{eq:omegaevent}
\end{equation}
where $\epsilon_\text{trig}$ is the trigger efficiency (Fig.~\ref{fig:Treff}) and $\epsilon_\text{vtx}$ is the primary vertex reconstruction
and selection efficiency, calculated with \PYTHIA{}6 as the ratio of the number of reconstructed events satisfying the
primary vertex selection to the total number of generated events.

The correction for the tracking efficiency and non-primary tracks, $\omega_\text{{trk}}(n_{\text{CMS}}, \pt,\eta )$, is defined as:
\ifthenelse{\boolean{cms@external}}{
\begin{multline}
 \omega_\text{{trk}}(n_{\text{CMS}}, \pt,\eta )  = \\
 \frac{1-f_{\text{np}}(n_{\text{CMS}},\pt,\eta)} {\epsilon_\text{{trk}}(n_{\text{CMS}},\pt,\eta) \left(1+f_{\mathrm{m}}(n_{\text{CMS}},\pt,\eta)\right)}.
\label{eq:omegatrack}
\end{multline}
}{
\begin{equation}
 \omega_\text{{trk}}(n_{\text{CMS}}, \pt,\eta )  = \frac{1-f_{\text{np}}(n_{\text{CMS}},\pt,\eta)} {\epsilon_\text{{trk}}(n_{\text{CMS}},\pt,\eta) \left(1+f_{\mathrm{m}}(n_{\text{CMS}},\pt,\eta)\right)}.
\label{eq:omegatrack}
\end{equation}
}
Here, $\epsilon_\text{{trk}}$ corrects for the geometric detector acceptance and the reconstruction tracking efficiency;
the correction factor $f_{\text{np}}$ accounts for the fraction of non-primary tracks, \ie secondary and misidentified
tracks, while $f_\mathrm{m}$ corrects for the effect of single charged particles that are
reconstructed multiple times.
These quantities are obtained from a detector simulation in bins of $n_\text{CMS}$, \pt, and $\eta$.  The effect of bin migrations is found to be
negligible. Reconstructed events are required to pass the event selection and the generated particles are matched to the reconstructed tracks
by using spatial and momentum information.

\begin{figure}[htb!]
  \begin{center}
\includegraphics[width=0.49\textwidth]{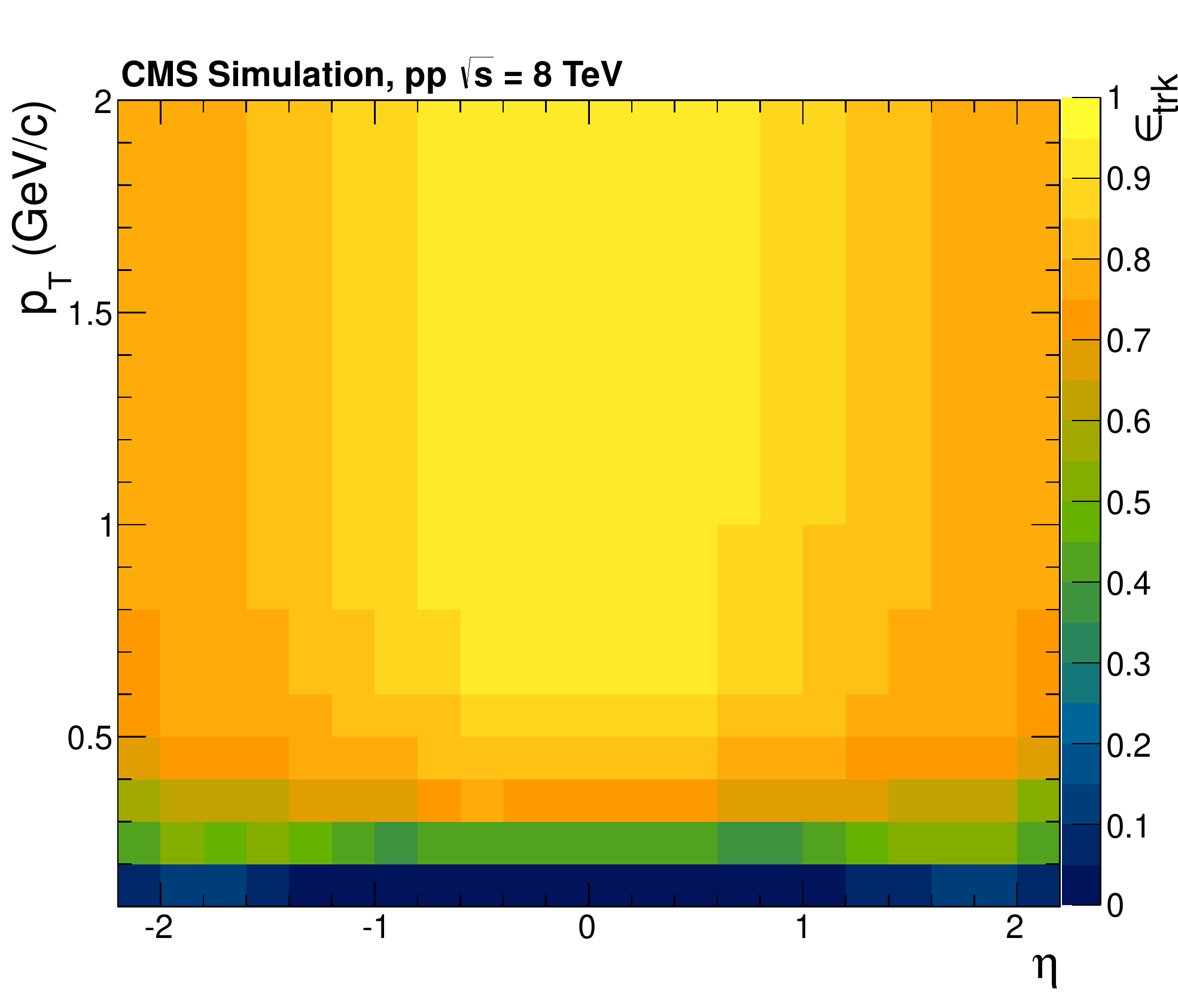}
\includegraphics[width=0.49\textwidth]{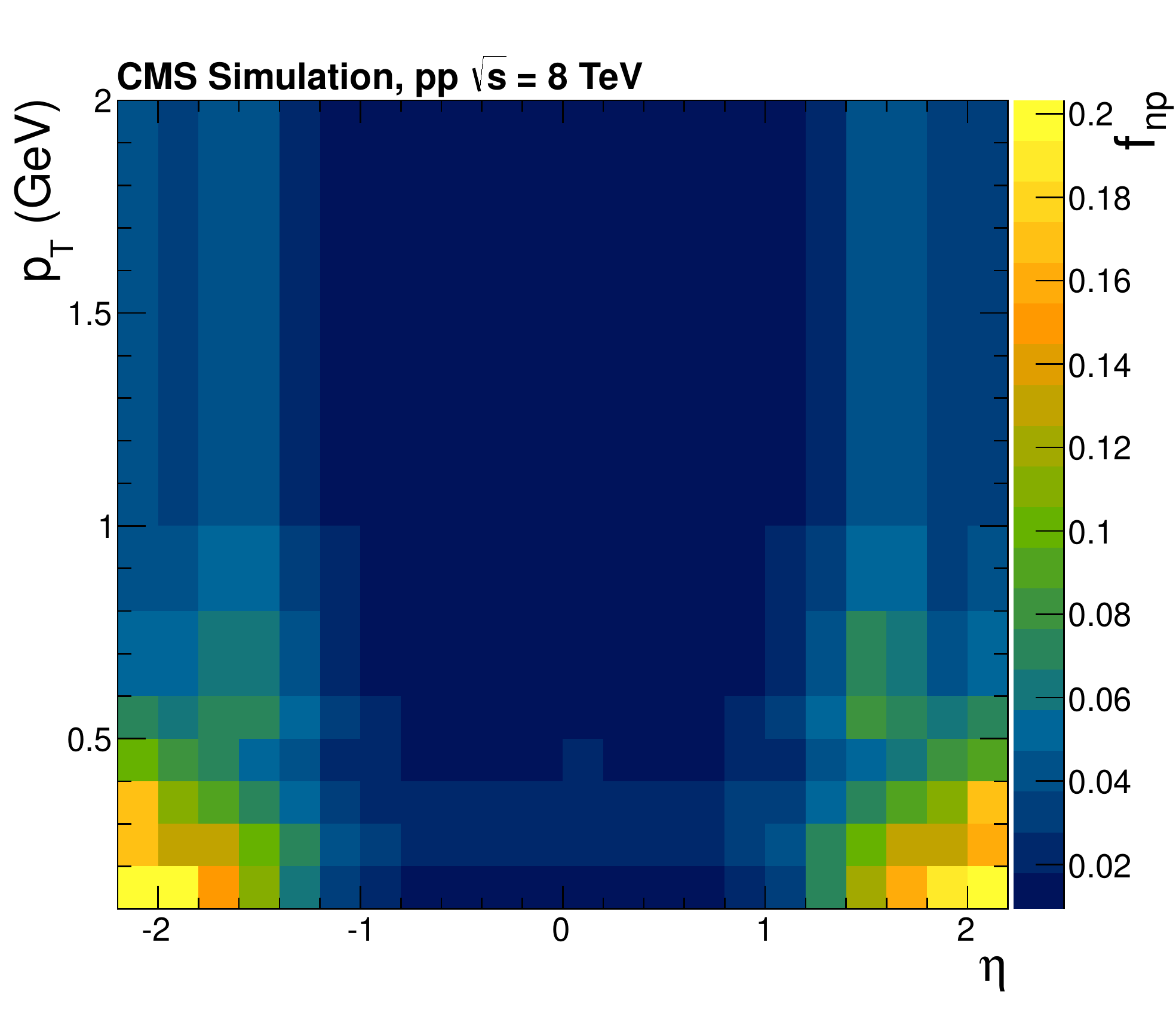}
    \caption{\cmsLeft: tracking efficiency, $\epsilon_\text{{trk}}$, as a function of \pt and $\eta$ and averaged over all
multiplicity bins ($n_{\text{CMS}}$), for tracks with $\pt>0.1$\GeVc and $\abs{\eta}<2.2$.
\cmsRight: correction factor, $f_{\mathrm{np}}$, for non-primary tracks as a function of \pt and
$\eta$ and averaged over all multiplicity bins ($n_{\text{CMS}}$), for tracks with $\pt>0.1$\GeVc and $\abs{\eta}<2.2$.}
    \label{fig:trackEff}
  \end{center}
\end{figure}
The tracking efficiency, $\epsilon_\text{{trk}}$, is determined as the ratio of the number
of all reconstructed tracks that are matched to generated particles and satisfy the track selection criteria
in an ($n_{\text{CMS}}, \pt,\eta$) bin to the number of generated primary charged particles in that bin.
As shown in Fig.~\ref{fig:trackEff} (\cmsLeft), $\epsilon_\text{{trk}}$ approaches unity for tracks with $\pt>0.5$\GeVc and $\abs{\eta}<1.5$.

The correction for non-primary tracks, $f_{\mathrm{np}}$, is estimated as the ratio of the number
of reconstructed  tracks not matched to a generated primary particle in a $n_{\text{CMS}}, \pt, \eta$ bin to all reconstructed
tracks in that bin.
The correction varies as a function of $\eta$ and \pt of the tracks,
 as shown in Fig.~\ref{fig:trackEff} (\cmsRight), and reaches its lowest values of about 2\%
for $\abs{\eta}<1.5$ and $\pt>0.5$\GeVc.
It becomes as large as 20\% at very low transverse momentum ($\pt<0.2$\GeVc) and large pseudorapidity ($\abs{\eta}>1.8$).

The correction factor for multiply reconstructed particles, $f_\mathrm{m}$, is estimated as the ratio of the number of generated primary charged particles that are associated
 to multiply reconstructed tracks in a given $n_{\text{CMS}}$, \pt, $\eta$ bin to the number of generated
charged particles in that bin. It is found to be below 1\%.

The model dependence of the corrections is determined by using \PYTHIA{}6, \EPOS, and \PYTHIA{}8.
The corrected data, based on correction factors derived from each generator independently, are found to differ by 1--4\% depending on the
pseudorapidity bin. This amount is taken as a systematic uncertainty.

The average correction factor for the event selection, $C_{\text{sel}}(\eta)$, defined in Section~\ref{sec:EventSelCorr_dndeta}, is found to be 1.01, 1.025, and 0.94 for the inclusive, NSD-enhanced, and SD-enhanced samples, respectively, independent of pseudorapidity.
The correction factor is obtained from \EPOS and \PYTHIA{}8. The average value of the correction factors from the
two generators is applied to the data, while the relative difference between the two generators
is taken as a systematic uncertainty.
In addition, the event selection bias and the corresponding systematic uncertainty is estimated for each $\eta$ bin as described in
Section~\ref{sec:EventSelCorr_dndeta} by comparing
the pseudorapidity distributions obtained with different methods. For the inclusive and NSD-enhanced samples,
the systematic uncertainty in
the event selection is found to be 3--5\% and 4--6\%, respectively, while for the SD-enhanced sample it is 9--16\%.

Corrections of 4--6\% are applied to the final results to extrapolate to \pt~=~0. The corrections are determined from the
$\rd{}N/\rd{}p_T$ spectrum of primary charged particles predicted by \PYTHIA{}6 and \PYTHIA{}8.
The corrections obtained from the two MC generators differ by about 3\%, resulting in a systematic uncertainty in
the $\rd{}N_{\text{ch}}/\rd{}\eta$ distributions of about 0.2\%. The same corrections are also estimated from
Tsallis fits~\cite{Tsallis} to the \pt distributions for each $\eta$ bin, giving consistent results.

A summary of the systematic uncertainties is given in Table~\ref{tab:sys}.
The uncertainties associated with the tracking efficiency are treated as uncorrelated between the $\eta$ bins.
For the inclusive and the NSD-enhanced samples, the most significant systematic uncertainties
are those due to the uncertainty in the tracking efficiency and the event selection.
The model dependence of $C_{\text{sel}}$ and
the uncertainty in the event selection are dominant for the SD-enhanced sample.
The total uncertainty in the tracking efficiency is estimated to be 3.9\% from a comparison of two-body
and four-body \PDz\ decays in data and simulated samples~\cite{trackingEfficiency}.
The uncertainties related to the primary vertex selection, the trigger efficiency, and pileup events
are found to be around 0.1\% and are neglected.
\begin{table*}
\centering
\topcaption{Systematic and statistical uncertainties of the $\rd{}N_{\textnormal{ch}}/\rd{}\eta$ measurement
in the central region. The given ranges indicate the $\eta$ dependence of the uncertainties.}
\label{tab:sys}
\begin{tabular}{lccc}
\hline
Source & Inclusive & NSD-enhanced & SD-enhanced \\
\hline
Event and primary track selection ($C_{\text{sel}}(\eta)$)  & 3--5\%  &  4--6\% & 9--16\% \\
Tracking efficiency          & 3.9\%   &  3.9\%  & 3.9\%    \\
Trigger efficiency           & 0.1\%   &  0.1\%  & 0.1\%    \\
Model dependence  of track corrections ($\omega_\text{{trk}}$)           & 1--4\%   &  1--4\%  & 1--4\%    \\
Correction to $\pt=0$        & 0.2\%   &  0.2\%  & 0.2\%    \\
\hline
Statistical                  & 0.1\%  & 0.1\% & 0.1\%  \\
\hline
Total                        & 5--7\% & 6--8\% & 10--17\%  \\ 
\hline
\end{tabular}
\end{table*}

\subsection{Measurement of \texorpdfstring{$\rd{}N_{\text{ch}}/\rd{}\eta$}{dN[ch]/d eta} in the forward region}
\label{sec:T2_dndeta}

The pseudorapidity density in the forward region is measured for each T2 half-arm independently, thus providing a
consistency check, as each half-arm differs in its alignment and track reconstruction efficiency.
The number of  primary track candidates passing the $z_\text{impact}$ parameter selection criteria is
calculated for each $\eta$ bin as a function
of the $z_\text{impact}$-value with the double-Gaussian and
exponential fits described in Section~\ref{sec:Event_selection}.
The fraction of tracks associated to the double-Gaussian distribution core ranges from about 74\% (lower $\abs{\eta}$ bins) to about
92\% (higher $\abs{\eta}$ bins),
and is used to weight each track by the probability for the track to be primary.

\begin{figure}[htb!]
\centering
\includegraphics[width=\cmsFigWidth]{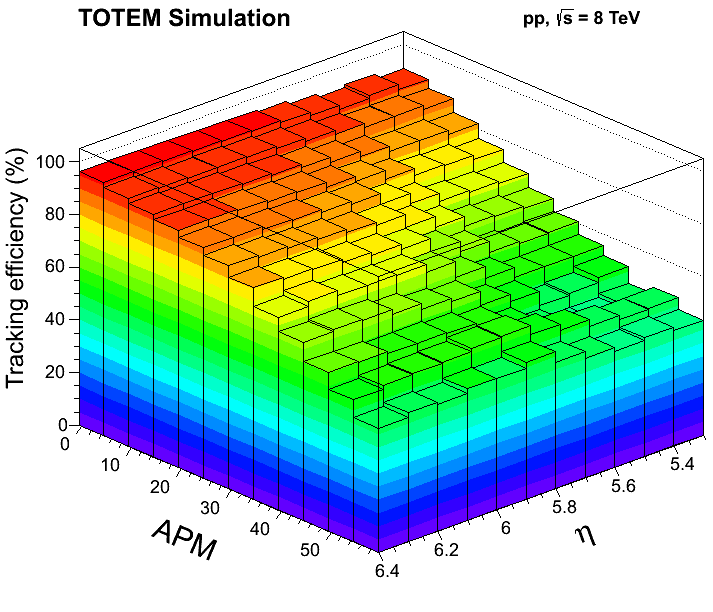}
\caption{Primary track efficiency as a function of $\eta$ and average pad cluster multiplicity
(APM) in one T2 half-arm for the inclusive pp sample. The efficiency includes the track primary-candidate condition.
Only particles with $\pt>40$\MeVc are considered.}
\label{fig:T2Effi}
\end{figure}

Each track is also weighted by the primary track efficiency, which depends on $\eta$ and on the average pad cluster
multiplicity (APM) per plane
in the corresponding half-arm. This efficiency, evaluated from MC generators, is defined as the probability to
successfully reconstruct, with a $z_\text{impact}$ parameter within the allowed region, a generated primary particle
that falls within the acceptance of the detector.
As shown in fig. 5 for one of the T2 half-arms, the tracking efficiency decreases with increasing APM,
since the reconstruction of tracks with sufficient number of hits becomes more difficult with larger occupancy.
The average primary track efficiency ranges from about 73\% (lower $\abs{\eta}$ bins) to about 87\% (higher $\abs{\eta}$ bins), as shown in Fig.~\ref{fig:T2Effi} for one of the T2 half-arms.
The APM probability is a rapidly decreasing distribution, with $\text{average}\sim24$ and $\textsc{rms} \sim29$, for the inclusive selection.
The rate of multiple associations of reconstructed tracks to the primary particle is negligible ($<$0.4\%) once the $z_\text{impact}$
requirement is imposed.

Conversion of photons from \Pgpz\ decays in the material between the IP and T2, as well as decay products of strange
particles, also contribute to the double-Gaussian peak.
The overall non-primary contribution,
to be subtracted from the double-Gaussian peak, was estimated as a function of $\eta$ with \PYTHIA{}8, \EPOS,
and \SIBYLL\ 2.1~\cite{Ahn:2009wx}.  The first two generators bracket the data in the forward region, while \SIBYLL is introduced to also enclose
the measurements from the LHCf experiment of the photon $\rd{}N/\rd{}E$ distribution~
\cite{lhcfgamma}. The value of this correction is about 18\% and
is obtained as the average of the three MC predictions.

{\tolerance=1000
The correction factors for the event selection bias ($C_{\text{sel}}(\eta)$) are about 1.05, 1.06, and 1.00
for the inclusive, NSD-enhanced, and SD-enhanced event samples, respectively.
\par}

Bin migration effects in $\eta$ are corrected for with \PYTHIA{}8, which gives the best description of the slope
of the measured $\rd{}N_{\textnormal{ch}}/\rd{}\eta$ distribution. The effects are typically at the few
percent level.

Events characterised by a high T2 hit multiplicity, typically due to showers generated by particles interacting with the material before T2, are not
included in the analysis. These events, for which track
reconstruction capability is limited, are characterised by an APM value larger than 60 and constitute 13.5\%,
16.5\%, and 6.3\% of the inclusive, NSD-enhanced, and
SD-enhanced samples, respectively.  The effect of removing these events is evaluated in a MC study, which yields
overall correction factors of about 1.05, 1.04, and 1.06 for the inclusive, NSD-enhanced, and
SD-enhanced samples, respectively. To verify the stability of this correction, the analysis is also repeated
after excluding events with APM values larger than 45 and re-evaluating the corresponding MC corrections.
The results of the two analyses agree within 1\%.
In addition, this correction is also estimated by extrapolating the measured average multiplicity obtained as a function of the
maximum APM included in the sample and without correcting for the missing fraction of the sample.
The extrapolation, performed with a second degree polynomial, gives a correction that is within the MC uncertainty.

The fully corrected $\rd{}N_{\text{ch}}/\rd{}\eta$ distributions in each $\eta$ bin are determined from:
\ifthenelse{\boolean{cms@external}}{
\begin{multline}
\label{s.long}
\frac{\rd{}N_\text{{ch}}}{\rd{} \eta}(\eta_i)  =\frac{2\pi}{\Delta\phi}\times\\
\frac{C_{\text{sel}}(\eta_i) \sum_{\text{evt}} \omega_{\text{evt}}(n_{\mathrm{T2}})\sum_{\text{trk} \in S_j}B_{ij}\,\omega_{\text{trk}}(\text{APM},\eta_j,z_\text{impact})}
{\Delta\eta\sum_{n_{\text{T2}}}N_{\text{evt}}(n_{\mathrm{T2}})\omega_{\text{evt}}
(n_{\mathrm{T2}})},
\end{multline}
}{
\begin{equation}
\label{s.long}
\frac{\rd{}N_\text{{ch}}}{\rd{} \eta}(\eta_i)  =
\frac{\,\,C_{\text{sel}}(\eta_i) \sum_{\text{evt}} \omega_{\text{evt}}(n_{\mathrm{T2}})\,\sum_{\text{trk} \in S_j}\,B_{ij}\,\omega_{\text{trk}}(\text{APM},\eta_j,z_\text{impact})}
{\Delta\eta\,\sum_{n_{\text{T2}}}N_{\text{evt}}(n_{\mathrm{T2}})\omega_{\text{evt}}
(n_{\mathrm{T2}})}\,\frac{2\pi}{\Delta\phi},
\end{equation}
}
where $S_j$ is the sample of tracks with $\eta_j-\Delta\eta/2 < \eta_j < \eta_j + \Delta\eta/2$ satisfying the selection criteria above, $\Delta\eta=0.05$ is the bin width, $C_{\text{sel}}$ is the correction factor related to the event selection defined in
Section~\ref{sec:EventSelCorr_dndeta}, $B_{ij}$ is the bin migration correction associated with the $j$th bin in $\eta$. In addition, $\Delta\phi/2\pi=192^{\circ}/360^{\circ}$ is the azimuthal acceptance of each T2 half-arm, $\omega_{\text{evt}}(n_{\mathrm{T2}})\equiv1/\epsilon_\text{trig}(n_{\mathrm{T2}})$ is the trigger efficiency correction, $N_{\text{evt}}(n_{\mathrm{T2}})$ is the number of selected events with track multiplicity $n_{\mathrm{T2}}$, and $\omega_{\text{trk}}$ is defined as:
\begin{equation}
\label{t2wtrk}
\omega_{\text{trk}}(\text{APM},\eta,z_\text{impact})=
\frac{P_\text{prim}(\eta,z_\text{impact})\,S_{\text{np}}(\eta)\,C_\text{mult}(\eta)}{\epsilon(\eta, \mathrm{APM})},
\end{equation}
where $P_\text{prim}$ is the probability for a track to be primary. Here, $\epsilon$ is the primary track efficiency,  $S_{\text{np}}$ is the correction
factor for the non-primary contribution to the double-Gaussian peak, and $C_{\text{mult}}$ is the correction factor for the exclusion of events
with APM values above 60.

The $\rd{}N_{\textnormal{ch}}/\rd{}\eta$ distribution thus obtained refers to charged particles
with $\pt>40$\MeVc, corresponding to the nominal $p_\text{T}$ acceptance of T2.
A MC-based estimation obtained with \EPOS LHC and \PYTHIA{}8 4C is used to correct the measurement
down to $\pt=0$. This correction, taken from the average of the two MC predictions, is about 2\%.

The evaluation of the systematic uncertainties for the $\rd{}N_{\textnormal{ch}}/\rd{}\eta$ distributions is performed similarly to that discussed in~\cite{TOTEM2012_dndeta}. Details are given in the following only for the most significant contributions.

The systematic uncertainty in the $P_\text{prim}$ function, which is of order 4--5\%, is evaluated
by including four effects: (a) the sensitivity to the misalignment corrections, quantified by varying the corrections
within their uncertainties, (b) the sensitivity to the $z_\text{impact}$ parameter fitting range, which was changed
by either one or two meters depending on the $\eta$ bin, (c) the sensitivity to the background parametrisation,
obtained by replacing the exponential function with a second degree polynomial, and (d) the difference between the area
estimated by the fitting function and the integral of the $z_{\text{impact}}$ distribution.

The systematic uncertainty in the primary-track efficiency is evaluated in studies where tracks are reconstructed with
a set of five detector planes (out of the total of ten) in a single T2 half-arm. The track reconstruction efficiency
is determined with the other set of detector planes in the same half-arm. The 5--6\% difference between the results
obtained from simulation and from data is taken as an estimate of the systematic uncertainty.

The systematic uncertainty due to non-primary tracks included in the double-Gaussian once the exponential contribution has been removed,
$S_{\text{np}}$, is evaluated by considering the range of the predictions of the \EPOS, \PYTHIA{}8,
and \SIBYLL\  MC generators, and is about 5\%.

The uncertainty in the correction for the exclusion of events with high secondary-particle multiplicity
($C_\text{mult}$) is taken as the difference between the \EPOS and \PYTHIA{}8 predictions,
which is about 3\%. The uncertainty on the correction for the event selection ($C_{\text{sel}}$)
is evaluated by taking into account both the dependence of the correction from the MCs mentioned above and the
dependence of the $\rd{}N_{\text{ch}}/\rd{}\eta$ results on the different event selection
strategies discussed earlier. This uncertainty is 13--15\% for the SD-enhanced sample and 2--3\% for the inclusive and NSD-enhanced samples.

The possible bias due to the material uncertainty and therefore on the production of secondary tracks is evaluated
as a function of $\eta$ from the MC vs. data discrepancy of the ratio between the number of tracks contained in the 96\%
double-Gaussian area and all the tracks in the same range. The average discrepancy is in the range of 2--6\%.
Simulation studies are also performed by varying the thickness of the material in front of T2 by 40\%. This part of the material
is the main source of secondary tracks that contribute to the double-Gaussian. The effect of the change of the material
results in a possible bias of less than 3\%.

Table~\ref{onebinsysttab} shows the uncertainties due to the corrections.
\begin{table*}
\centering
\topcaption{Systematic and statistical uncertainties of the $\rd{}N_{\textnormal{ch}}/\rd{}\eta$ measurement with the
T2 detector in the forward region. The first two contributions are half-arm dependent and partly $\eta$-uncorrelated, while the remaining, excluding the statistical one, are half-arm independent and correlated across bins in $\eta$. The given ranges indicate the $\eta$ dependence of the uncertainties.}
\label{onebinsysttab}
\begin{tabular}{lccc}
\multicolumn{2}{c}{} \\
\hline
  \multicolumn{1}{c}{Source} & Inclusive & NSD-enhanced  & SD-enhanced   \\
  \hline
Tracking efficiency data-MC discrepancy          &  5--6\%  &  5--6\% &   5--6\% \\
Primary selection (including alignment)        &   4--5\%   &   4--5\%   &   4--5\%   \\
Non-primaries in the double-Gaussian peak              &   5\%   &  5\%   &   5\%   \\
Material effects                               &   3--6\%   &  3--6\%   &   3--6\%   \\
High-multiplicity events          &   3\%   &   3\%   &  3\% \\
Event selection                   &  2--3\% &  2--3\% &  13--15\% \\
Tracking efficiency dependence      &   2\%   &   2\%   &   2\%   \\
~~on energy spectrum and  magnetic field                      &  &  &  \\
Track quality criterion                        &   1\%   &   1\%   &   1\%   \\
Correction to $\pt=0$                          &   0.5\%   &   0.5\%   &   0.5\%   \\
Trigger efficiency                                   &   0.2\%   &   0.2\%   &   0.2\%   \\
\hline
Statistical                                   &   0.1\%   &   0.1\%   &   0.1\%   \\
\hline
Total (after averaging half-arms)                &   10--12\%   &   10--12\%   &   16--18\% \\
\hline
\end{tabular}
\end{table*}
To compute the total systematic uncertainty the errors are first separated into half-arm-correlated and uncorrelated
parts and a weighted average between the four half-arms is taken. The $\rd{}N_{\textnormal{ch}}/\rd{}\eta$ measurements obtained for the different T2 half-arms are found to be compatible.

The first two systematic uncertainties in Table~\ref{onebinsysttab} vary as a function of $\eta$ and contribute
to the uncorrelated bin-by-bin uncertainties. Conversely, the remaining systematic uncertainties affect all $\eta$ bins in the same direction.
The effect that systematic uncertainties might introduce in the difference of the $\rd{}N_{\textnormal{ch}}/\rd{}\eta$
values at the beginning and at the end of the T2 acceptance is estimated to be at most 7\%.

The total uncertainty is obtained by adding in
quadrature the $\eta$-correlated uncertainty and the $\eta$-uncorrelated one and the (negligible)
statistical uncertainty.

\section{Results}
\begin{figure*}[htb!p]
\centering
\includegraphics[width=0.45\textwidth]{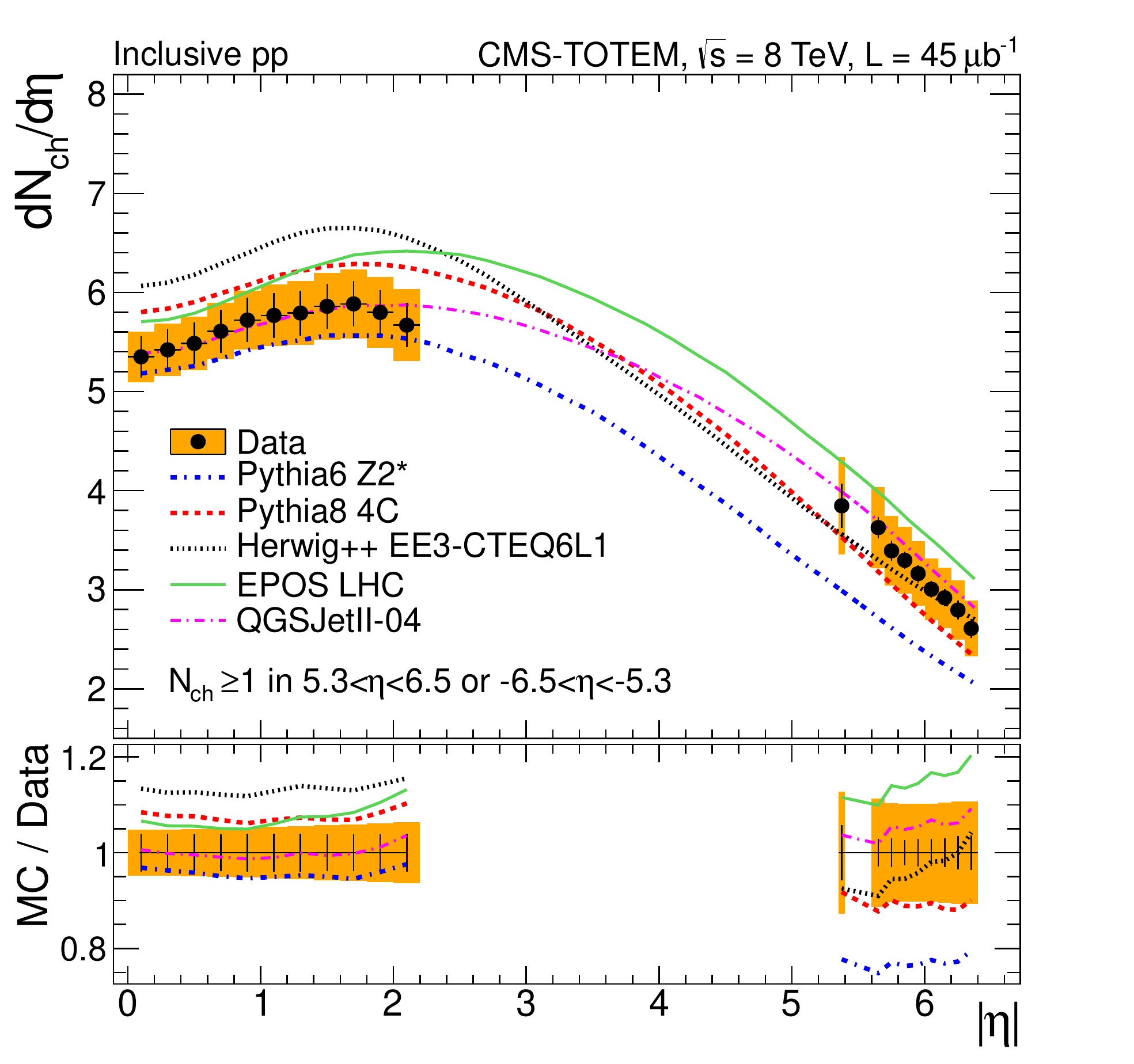}
\includegraphics[width=0.45\textwidth]{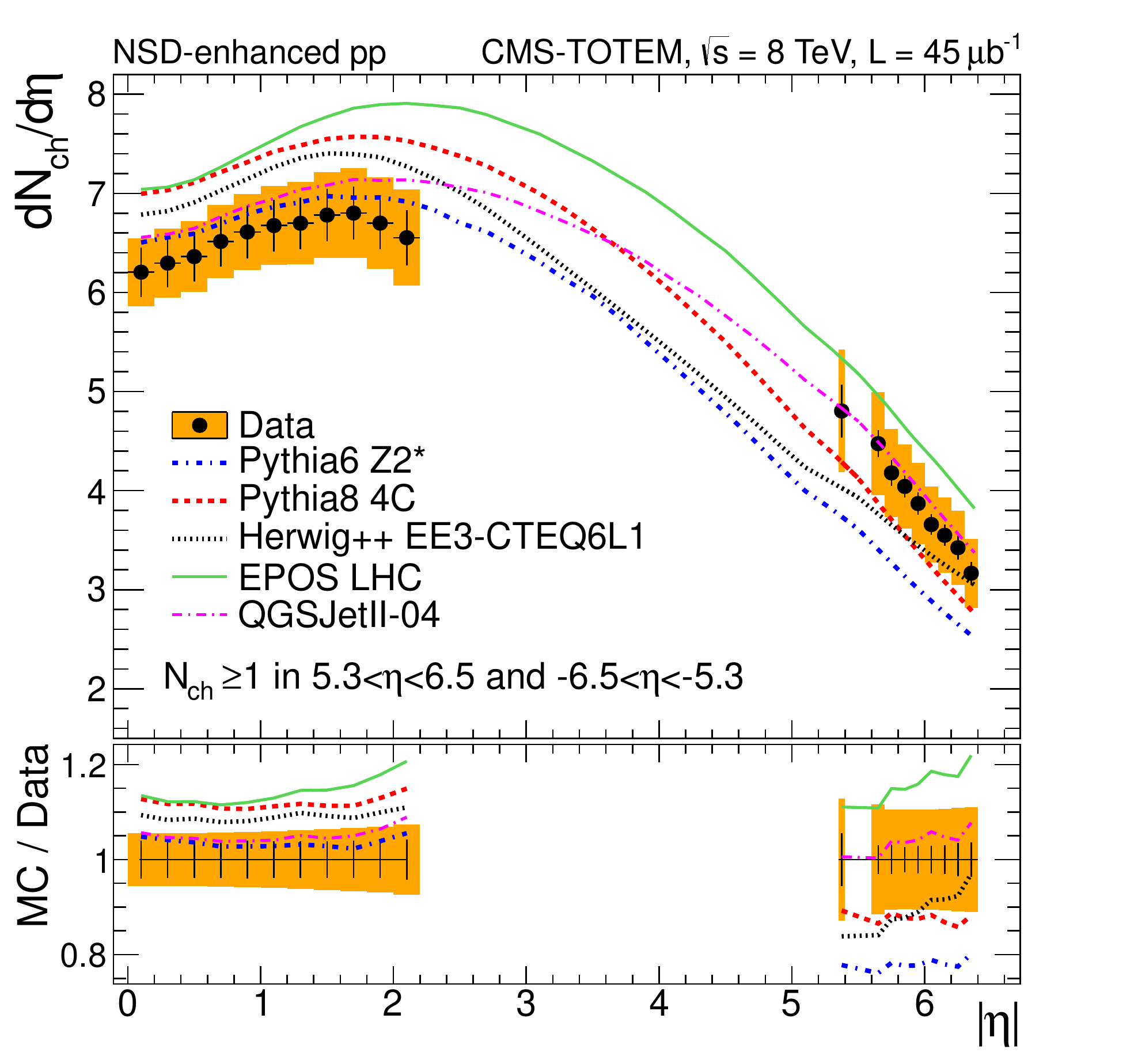}
\includegraphics[width=0.50\textwidth ]{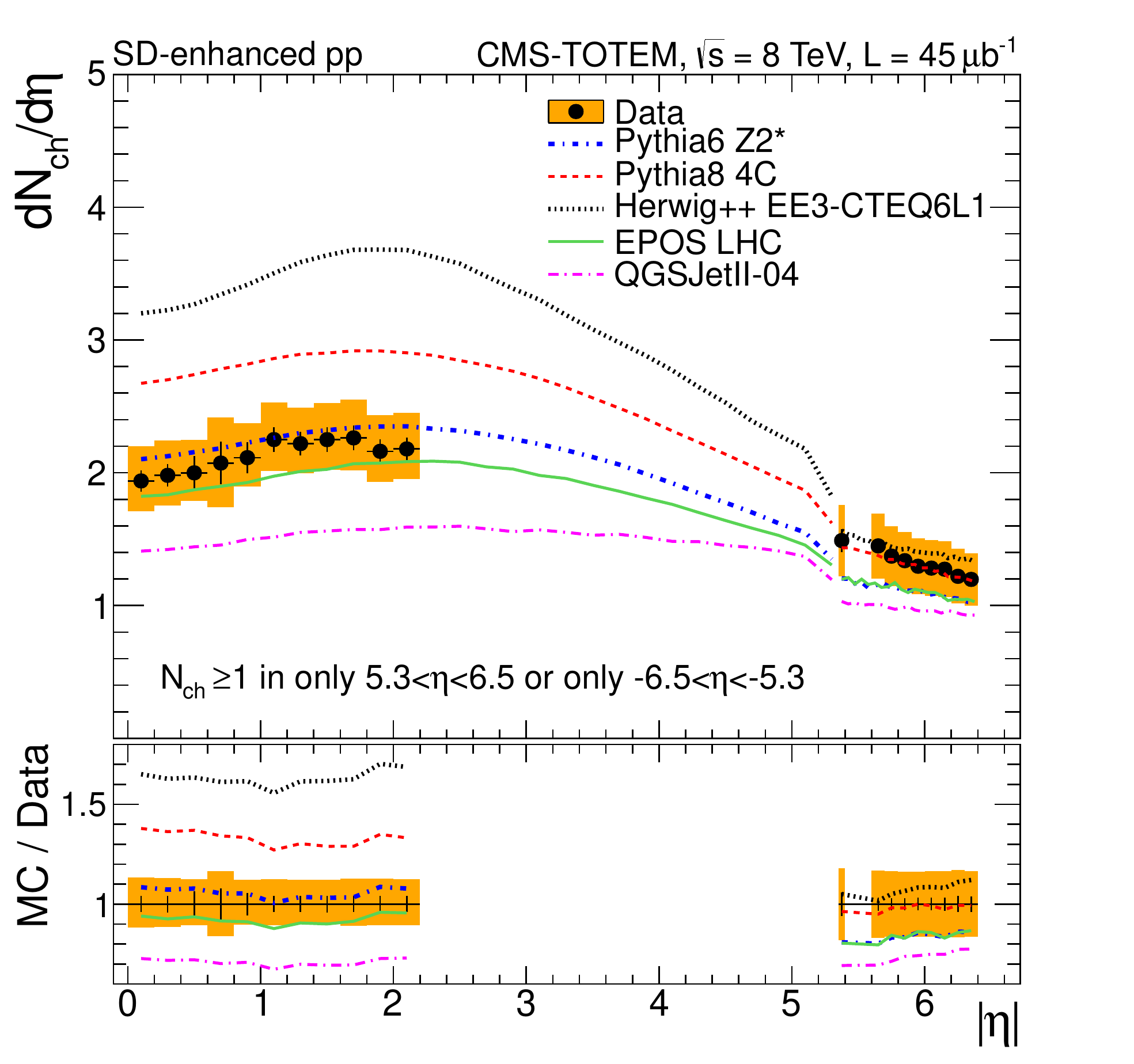}
\caption{
Charged-particle pseudorapidity distributions from an inclusive sample (top left), a NSD-enhanced sample (top right),
and a SD-enhanced sample (bottom). The error bars represent the statistical $+$ uncorrelated systematics
between neighbouring bins and
the bands show the combined systematic and statistical uncertainties.
The measurements are compared to results from \PYTHIA{}6, tune Z2*, \PYTHIA{}8, tune 4C, \HERWIGpp, tune UE-EE-3 with
CTEQ6L1 PDFs, \EPOS, tune LHC, and \QGS-04.}
 \label{fig:datacorr}
\end{figure*}

The combined CMS-TOTEM charged particle pseudorapidity distributions are presented in Fig.~\ref{fig:datacorr}
for the three event selection samples shown in Table~\ref{tab_selection}. The results are derived in the central
region by averaging the data points in the
corresponding $\pm\eta$ bins and in the forward region by averaging over the four T2 half-arms. The uncertainty band represents
the total uncertainty, while the bars show the statistical and uncorrelated systematics
between neighbouring bins.

\begin{table*}[!hbHp]
\centering
\topcaption{Event selection criteria applied at the stable-particle level in the MC simulation.
\label{tab_selection}}
\begin{tabular}{lllll}
\hline
\multicolumn{5}{l}{Inclusive sample}  \\
\quad$N_{\text{charged particles}} >0 $ &in&  $5.3 <\eta< 6.5$ & or & $-6.5<\eta<-5.3$, $\pt>0$     \\
\multicolumn{5}{l}{NSD-enhanced sample} \\
\quad$N_{\text{charged particles}} >0 $ &in&  $5.3 <\eta< 6.5$ &and & $-6.5<\eta<-5.3$, $\pt>0$     \\
\multicolumn{5}{l}{SD-enhanced sample} \\
\quad$N_{\text{charged particles}} >0 $ &in only& $5.3 <\eta< 6.5$ &or only in & $-6.5<\eta<-5.3$, $\pt>0$    \\
\hline
\end{tabular}
\end{table*}

In the central region, the pseudorapidity density at $\eta=0$ is $5.35 \pm 0.36$ for the inclusive
sample, $6.20 \pm 0.46$  for the NSD-enhanced sample, and $1.94^{\:+\:0.26}_{\:-\:0.23}$ for the SD-enhanced sample, with
negligible statistical uncertainties.
The predictions from various MC event generators differ from the data by up to 20\% for the
inclusive and NSD-enhanced samples, with even larger discrepancies for the SD-enhanced sample.
The data are well described by \PYTHIA{}6 and \QGS-04 for the inclusive selection. For the NSD-enhanced sample,
the predictions obtained from \PYTHIA{}6 and \QGS-04 agree with the data for most
$\eta$ bins.
A good description of the measurement for the SD-enhanced sample is provided by both \EPOS and \PYTHIA{}6.

The forward pseudorapidity density decreases with $\abs{\eta}$.
In the inclusive sample, $\rd{}N_{\text{ch}}/\rd{}\eta$ is $3.85 \pm 0.49$ at $\eta=5.375$ and $2.61 \pm 0.28$ at $\eta=6.350$, with negligible statistical uncertainty. The  pseudorapidity density of the NSD-enhanced sample ranges between $4.80 \pm 0.62$ and $3.17 \pm 0.35$, while for the SD-enhanced sample it is in the range of $1.49 \pm 0.27$ to $1.20 \pm 0.20$.
The MC predictions for the three samples differ from the data by up to about $\pm$30\%. For the inclusive and NSD-enhanced samples, the data in the forward region are in agreement with the prediction from \QGS-04 and are between the \EPOS and \PYTHIA{}8 results.
For the SD-enhanced selection, the TOTEM data points are close to the \PYTHIA{}8 and \HERWIGpp predictions, while \QGS-04 underestimates the data.
The change in the slope of the MC curves close to $\eta=5.3$, more visible for the NSD- and SD-enhanced distributions, is due to the event
selection requirement of at least one charged particle in the pseudorapidity region of T2.

\begin{figure}[htb!]
\centering
\includegraphics[width=\cmsFigWidth]{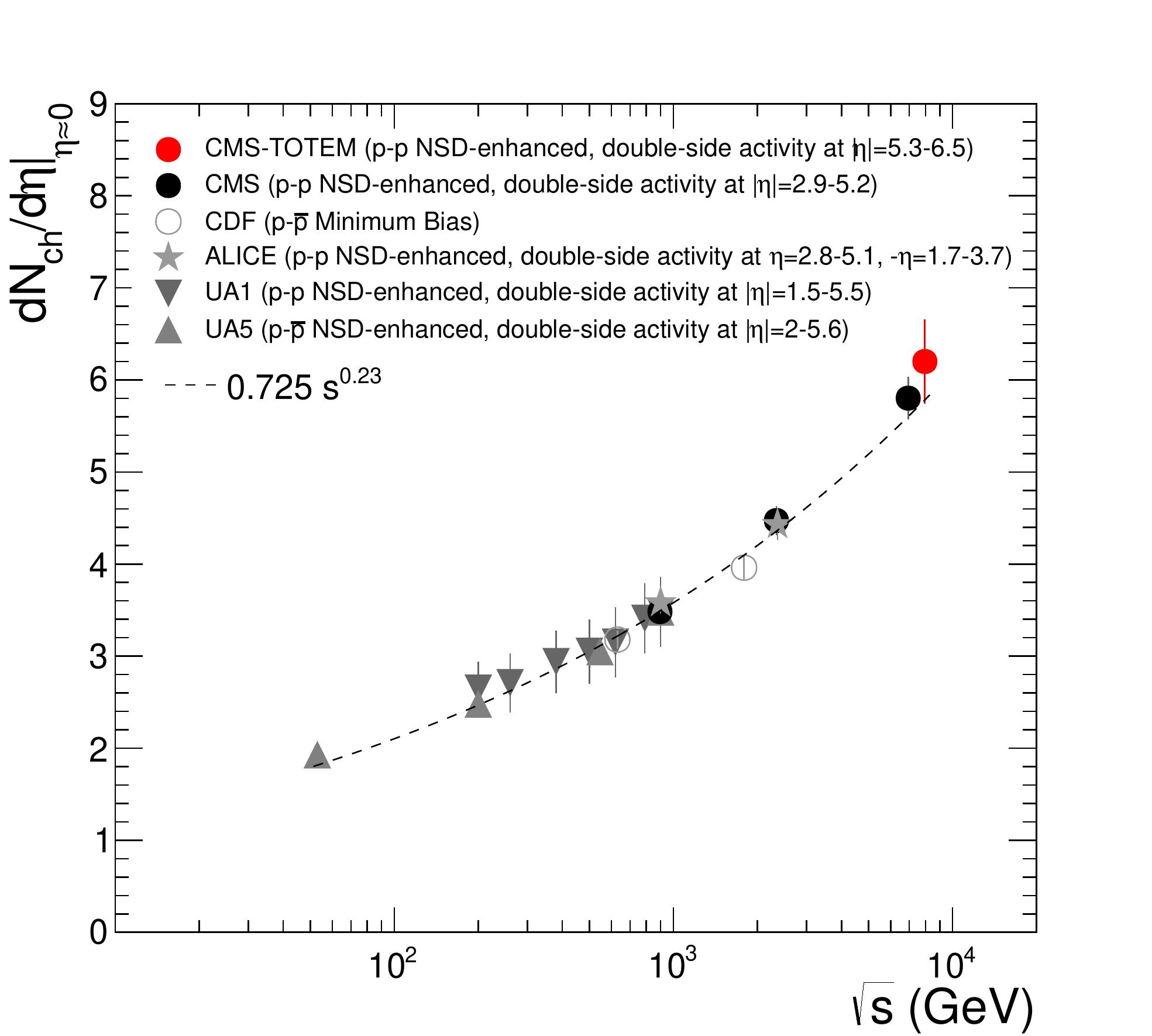}
  \caption{Value of $\rd{}N_{\text{ch}}/\rd{}\eta$ at $\eta\approx0$ as a function of the centre-of-mass energy in pp and
   \Pp\Pap\ collisions. Shown are measurements performed with different NSD event selections from
UA1~\cite{UA1}, UA5~\cite{UA5}, CDF~\cite{CDF:2009, CDF:1990}, ALICE~\cite{ALICE2010a}, and CMS~\cite{CMSdndeta2}.
The dashed line is a power-law fit to the data.}
    \label{fig:sqrts}
\end{figure}

The centre-of-mass energy dependence of the pseudorapidity distribution at $\eta\approx0$ is shown in Fig.~\ref{fig:sqrts}, which includes
data from various other experiments for NSD events in pp and p\={p} collisions. Although the different experiments
do not use identical event selection criteria, they all include a large fraction of NSD events.
Particle production at $\eta\approx0$ is
expected to follow a power-law dependence, $\rd{}N_\text{ch}/\rd{}\eta\left|_{\eta=0}\right. \propto s^\epsilon$, with
$\epsilon$ in the range 0.14--0.24~\cite{denterria:2011}.
The result of fitting the high-energy pp and $\Pp\Pap$ central-pseudorapidity particle densities with this function is shown in Fig.~\ref{fig:sqrts}. A value of $\epsilon = 0.23 \pm 0.01$ is obtained.

\section{Summary}
Measurements of charged-particle densities over a large pseudorapidity range are presented
for proton-proton (pp) collisions at a centre-of-mass energy of 8\TeV. The data were collected concurrently with
the CMS and TOTEM detectors during a dedicated run with low probability for overlapping pp interactions in the same
bunch crossing, and correspond to an integrated luminosity of $\mathcal{L} = 45\mubinv$.
Pseudorapidity distributions of charged particles within $\abs{\eta} < 2.2$ and $5.3 < \abs{\eta} < 6.4$ have been measured for three event samples with different final state topologies: a sample of inclusive inelastic pp events, a sample dominated by non single diffractive dissociation (NSD) events, and a sample enriched in single diffractive dissociation (SD) events. The data are compared to theoretical predictions obtained from five different MC event generators and tunes (\PYTHIA{}6 Z2*, \PYTHIA{}8 4C, \HERWIGpp UE-EE-3, \EPOS LHC tune, and \QGS-04).

In the central region, the inclusive and NSD-enhanced samples are well described by \PYTHIA{}6 and \QGS.  For the SD-enhanced sample a good description of the data is provided by both \PYTHIA{}6 and \EPOS. In the forward region, the pseudorapidity distributions for the inclusive and NSD-enhanced samples are between the \PYTHIA{}8 and \EPOS predictions.  The \QGS predictions are compatible with the data.  The pseudorapidity distribution in the SD-enhanced sample, affected by a larger systematic uncertainty, is best described by \PYTHIA{}8 and \HERWIGpp.

The charged-particle densities obtained in this paper span the largest pseudorapidity interval ever measured at the LHC and have
the unique potential to probe the correlation between particle production in the central region and
that in the forward region. With the tunes used, none of the MC event generators are able to consistently describe the data
over the whole $\eta$ region and for all event samples.

\section*{Acknowledgements}
\hyphenation{Bundes-ministerium Forschungs-gemeinschaft Forschungs-zentren}
{\tolerance=1200
We congratulate our colleagues in the CERN accelerator departments for the excellent performance of the LHC. We especially thank the beam optics development team for the design and the successful commissioning of the high $\beta$-optics and the LHC machine coordinators for scheduling the dedicated fills. We are grateful to  the technical and administrative staffs at CERN and at other CMS and TOTEM institutes for their contributions to the success of the CMS and TOTEM efforts. In addition, we gratefully acknowledge the computing centres and personnel of the Worldwide LHC Computing Grid for delivering so effectively the computing infrastructure essential to our analyses.
We acknowledge the enduring support for the construction and operation of the LHC and the CMS and TOTEM detectors provided by our affiliated institutions as included in the lists of authors and the following funding agencies: the Austrian Federal Ministry of Science, Research and Economy and the Austrian Science Fund; the Belgian Fonds de la Recherche Scientifique, and Fonds voor Wetenschappelijk Onderzoek; the Brazilian Funding Agencies (CNPq, CAPES, FAPERJ, and FAPESP); the Bulgarian Ministry of Education, Youth and Science; CERN; the Chinese Academy of Sciences, Ministry of Science and Technology, and National Natural Science Foundation of China; the Colombian Funding Agency (COLCIENCIAS); the Croatian Ministry of Science, Education and Sport, and the Croatian Science Foundation; the Research Promotion Foundation, Cyprus; the Ministry of Education and Research, Estonian Research Council via IUT23-4 and IUT23-6 and European Regional Development Fund, Estonia; the Academy of Finland, Finnish Ministry of Education and Culture, and Helsinki Institute of Physics; the Institut National de Physique Nucl\'eaire et de Physique des Particules~/~CNRS, and Commissariat \`a l'\'Energie Atomique et aux \'Energies Alternatives~/~CEA, France; the Bundesministerium f\"ur Bildung und Forschung, Deutsche Forschungsgemeinschaft, and Helmholtz-Gemeinschaft Deutscher Forschungszentren, Germany; the General Secretariat for Research and Technology, Greece; the National Scientific Research Foundation, National Innovation Office, and the OTKA grant NK 101438, 73143, Hungary; the Department of Atomic Energy and the Department of Science and Technology, India; the Institute for Studies in Theoretical Physics and Mathematics, Iran; the Science Foundation, Ireland; the Istituto Nazionale di Fisica Nucleare, Italy; the Korean Ministry of Education, Science and Technology and the World Class University program of NRF, Republic of Korea; the Lithuanian Academy of Sciences; the Ministry of Education, and University of Malaya (Malaysia); the Mexican Funding Agencies (CINVESTAV, CONACYT, SEP, and UASLP-FAI); the Ministry of Business, Innovation and Employment, New Zealand; the Pakistan Atomic Energy Commission; the Ministry of Science and Higher Education and the National Science Centre, Poland; the Funda\c{c}\~ao para a Ci\^encia e a Tecnologia, Portugal; JINR (Armenia, Belarus, Georgia, Ukraine, Uzbekistan); the Ministry of Education and Science of the Russian Federation, the Federal Agency of Atomic Energy of the Russian Federation, Russian Academy of Sciences, and the Russian Foundation for Basic Research; the Ministry of Education, Science and Technological Development of Serbia; the Secretar\'{\i}a de Estado de Investigaci\'on, Desarrollo e Innovaci\'on and Programa Consolider-Ingenio 2010, Spain; the Swiss Funding Agencies (ETH Board, ETH Zurich, PSI, SNF, UniZH, Canton Zurich, and SER); the Ministry of Science and Technology, Taipei; the Thailand Center of Excellence in Physics, the Institute for the Promotion of Teaching Science and Technology of Thailand, Special Task Force for Activating Research and the National Science and Technology Development Agency of Thailand; the Scientific and Technical Research Council of Turkey, and Turkish Atomic Energy Authority; the National Academy of Sciences of Ukraine, and State Fund for Fundamental Researches, Ukraine; the Science and Technology Facilities Council, UK; the US Department of Energy, and the US National Science Foundation.

Individuals have received support from the Marie-Curie programme and the European Research Council and EPLANET (European Union); the Leventis Foundation; the A. P. Sloan Foundation; the Alexander von Humboldt Foundation; the Belgian Federal Science Policy Office; the Fonds pour la Formation \`a la Recherche dans l'Industrie et dans l'Agriculture (FRIA-Belgium); the Agentschap voor Innovatie door Wetenschap en Technologie (IWT-Belgium); the Ministry of Education, Youth and Sports (MEYS) of the Czech Republic; the Magnus Ehrnrooth foundation and the Waldemar von Frenckell foundation, Finland; the Finnish Academy of Science and Letters (The Vilho, Yrj{\"o} and Kalle V{\"a}is{\"a}l{\"a} Fund), and the Charles Simonyi Fund (Hungary); the Council of Science and Industrial Research, India; the Compagnia di San Paolo (Torino); the HOMING PLUS programme of Foundation for Polish Science, cofinanced by EU, Regional Development Fund; and the Thalis and Aristeia programmes cofinanced by EU-ESF and the Greek NSRF.
\par}

\bibliography{auto_generated}   
\ifthenelse{\boolean{cms@external}}{}{
\ifthenelse{\boolean{cms@final}}{
\clearpage
\appendix
\section{The CMS Collaboration \label{app:collab}}\begin{sloppypar}\hyphenpenalty=5000\widowpenalty=500\clubpenalty=5000\textbf{Yerevan Physics Institute,  Yerevan,  Armenia}\\*[0pt]
S.~Chatrchyan, V.~Khachatryan, A.M.~Sirunyan, A.~Tumasyan
\vskip\cmsinstskip
\textbf{Institut f\"{u}r Hochenergiephysik der OeAW,  Wien,  Austria}\\*[0pt]
W.~Adam, T.~Bergauer, M.~Dragicevic, J.~Er\"{o}, C.~Fabjan\cmsAuthorMark{1}, M.~Friedl, R.~Fr\"{u}hwirth\cmsAuthorMark{1}, V.M.~Ghete, C.~Hartl, N.~H\"{o}rmann, J.~Hrubec, M.~Jeitler\cmsAuthorMark{1}, W.~Kiesenhofer, V.~Kn\"{u}nz, M.~Krammer\cmsAuthorMark{1}, I.~Kr\"{a}tschmer, D.~Liko, I.~Mikulec, D.~Rabady\cmsAuthorMark{2}, B.~Rahbaran, H.~Rohringer, R.~Sch\"{o}fbeck, J.~Strauss, A.~Taurok, W.~Treberer-Treberspurg, W.~Waltenberger, C.-E.~Wulz\cmsAuthorMark{1}
% ...
\vskip\cmsinstskip
% ...
1:~~Also at Vienna University of Technology, Vienna, Austria\\
2:~~Also at CERN, European Organization for Nuclear Research, Geneva, Switzerland\\
% ...

\clearpage\section{The TOTEM Collaboration}

G.~Antchev,$^{15}$
P.~Aspell,$^{8}$
I.~Atanassov,$^{8,a}$
V.~Avati,$^{8}$
J.~Baechler,$^{8}$
V.~Berardi,$^{5a,5b}$
M.~Berretti,$^{7b}$
E.~Bossini,$^{7b}$
U.~Bottigli,$^{7b}$
M.~Bozzo,$^{6a,6b}$
E.~Br\"{u}cken,$^{3a,3b}$
A.~Buzzo,$^{6b}$
F.~S.~Cafagna,$^{5a}$
M.~G.~Catanesi,$^{5a}$
C.~Covault,$^{9}$
M.~Csan\'{a}d,$^{4,d}$
T.~Cs\"{o}rg\H{o},$^{4}$
M.~Deile,$^{8}$
M.~Doubek,$^{1b}$
K.~Eggert,$^{9}$
V.~Eremin,$^{12}$
A. Fiergolski,$^{5a,b}$
F.~Garcia,$^{3a}$
V.~Georgiev,$^{11}$
S.~Giani,$^{8}$
L.~Grzanka,$^{10,c}$
J.~Hammerbauer,$^{11}$
J.~Heino,$^{3a}$
T.~Hilden,$^{3a,3b}$
A.~Karev,$^{8}$
J.~Ka\v{s}par,$^{1a,8}$
J.~Kopal,$^{1a,8}$
J.~Kosinski$^{10}$
V.~Kundr\'{a}t,$^{1a}$
S.~Lami,$^{7a}$
G.~Latino,$^{7b}$
R.~Lauhakangas,$^{3a}$
T.~Leszko,$^{16}$
E.~Lippmaa,$^{2}$
J.~Lippmaa,$^{2}$
M.~V.~Lokaj\'{i}\v{c}ek,$^{1a}$
L.~Losurdo,$^{7b}$
F.~Lucas~Rodr\'{i}guez,$^{8}$
M.~Macr\'{i},$^{6b}$
T.~M\"aki,$^{3a}$
A.~Mercadante,$^{5a}$
N.~Minafra,$^{5b,8}$
S.~Minutoli,$^{6b}$
F.~Nemes,$^{4,d}$
H.~Niewiadomski,$^{8}$
E.~Oliveri,$^{7b}$
F.~Oljemark,$^{3a,3b}$
R.~Orava,$^{3a,3b}$
M.~Oriunno,$^{13}$
K.~\"{O}sterberg,$^{3a,3b}$
P.~Palazzi,$^{7b}$
Z.~Peroutka,$^{11}$
J.~Proch\'{a}zka,$^{1a}$
M.~Quinto,$^{5a,5b}$
E.~Radermacher,$^{8}$
E.~Radicioni,$^{5a}$
F.~Ravotti,$^{8}$
L.~Ropelewski,$^{8}$
G.~Ruggiero,$^{8}$
H.~Saarikko,$^{3a,3b}$
A.~Scribano,$^{7b}$
J.~Smajek,$^{8}$
W.~Snoeys,$^{8}$
J.~Sziklai,$^{4}$
C.~Taylor,$^{9}$
N.~Turini,$^{7b}$
V.~Vacek,$^{1b}$
J.~Welti,$^{3a,3b}$
J.~Whitmore,$^{14}$
P.~Wyszkowski,$^{10}$
K.~Zielinski$^{10}$

\vskip 0pt plus 4pt
$^{ 1a }$ Institute of Physics of the Academy of Sciences of the Czech Republic, Praha, Czech Republic\\
$^{ 1b }$ Czech Technical University, Praha, Czech Republic\\
$^{ 2 }$ National Institute of Chemical Physics and Biophysics NICPB, Tallinn, Estonia\\
$^{ 3a }$ Helsinki Institute of Physics, Helsinki, Finland\\
$^{ 3b }$ Department of Physics,  University of Helsinki, Helsinki, Finland\\
$^{ 4 }$ MTA Wigner Research Center,  RMKI, Budapest, Hungary\\
$^{ 5a }$ INFN Sezione di Bari, Bari, Italy\\
$^{ 5b }$ Dipartimento Interateneo di Fisica di  Bari, Bari, Italy\\
$^{ 6a }$ Universit\`{a} degli Studi di Genova,  Genova, Italy\\
$^{ 6b }$ INFN Sezione di Genova, Genova, Italy\\
$^{ 7a }$ INFN Sezione di Pisa, Pisa, Italy\\
$^{ 7b }$ Universit\`{a} degli Studi di Siena and Gruppo Collegato INFN di Siena,  Siena, Italy\\
$^{ 8 }$ CERN,   Geneva, Switzerland\\
$^{ 9 }$ Case Western Reserve University,  Dept of Physics, Cleveland, OH, USA\\
$^{ 10 }$ AGH University of Science and Technology, Krakow, Poland\\
$^{ 11 }$ University of West Bohemia, Pilsen, Czech Republic\\
$^{12}$ Ioffe Physical -- Technical Institute of Russian Academy of Sciences, StPetersburg, Russia\\
$^{13}$ SLAC National Accelerator Laboratory, Stanford, CA, USA\\
$^{14}$  Penn State University, Dept~of Physics, University Park, PA USA\\
$^{15}$   INRNE-BAS, Institute for Nuclear Research and Nuclear Energy, Bulgarian Academy of Sciences, Sofia, Bulgaria\\
$^{16}$ Warsaw University of Technology, Warsaw, Poland\\
\vskip 0pt plus 4pt
$^a$ Also at INRNE-BAS, Institute for Nuclear Research and Nuclear Energy, Bulgarian Academy of Sciences, Sofia, Bulgaria\\
$^b$ Also at Warsaw University of Technology, Warsaw, Poland\\
$^c$ Also at  Institute of Nuclear Physics, Polish Academy of Science, Cracow, Poland\\
$^d$ Also at Department of Atomic Physics, E\"otv\"os University,  Budapest, Hungary\\

\end{sloppypar}}{}
}
\end{document}